\documentclass[
reprint,
superscriptaddress,
amsmath,
amssymb,
pra,
floatfix,
twocolumn,
]{revtex4-1}
\usepackage{float}
\usepackage{amsfonts,amsmath,amssymb,amsthm,mathtools}
\usepackage{braket}
 \usepackage{array,bm,color}
\usepackage{epsfig,graphicx,nomencl,revsymb4-1,upgreek,url}
\usepackage{hyperref}
\usepackage{caption}
\usepackage{subcaption}
\usepackage{graphicx}
\usepackage{calc}
\usepackage{siunitx}
\newcolumntype{Y}{>{\centering\arraybackslash}X}
\graphicspath{{./figures/}}

\newcommand{\note}[1]{}

\usepackage[utf8]{inputenc}

\begin{document}
\author{Anthony M. Polloreno}
\affiliation{JILA, NIST and Department of Physics, University of Colorado, 440 UCB, Boulder, Colorado 80309, USA}
\email[Email: ]{ampolloreno@gmail.com}
\author{Ana Maria Rey}
\affiliation{JILA, NIST and Department of Physics, University of Colorado, 440 UCB, Boulder, Colorado 80309, USA}
\affiliation{Center for Theory of Quantum Matter, University of Colorado, Boulder, CO 80309, USA}
\author{John J. Bollinger}
\affiliation{National Institute of Standards and Technology, Boulder, Colorado 80305, USA}
\date{\today}
\title{Individual qubit addressing of rotating ion crystals in a Penning trap}

\begin{abstract}
Trapped ions boast long coherence times and excellent gate fidelities, making them a useful platform for quantum information processing. Scaling to larger numbers of ion qubits in RF Paul traps demands great effort. Another technique for trapping ions is via a Penning trap where a 2D crystal of hundreds of ions is formed by controlling the rotation of the ions in the presence of a strong magnetic field. However, the rotation of the ion crystal makes single ion addressability a significant challenge. We propose a protocol that takes advantage of a deformable mirror to introduce AC Stark shift patterns that are static in the rotating frame of the crystal.  Through numerical simulations we validate the potential of this protocol to perform high-fidelity single-ion gates in crystalline arrays of hundreds of ions.

\end{abstract}

\maketitle
\section{Introduction}
\begin{figure*}
    \centering
         \begin{subfigure}[t]{0.3\textwidth}
    \centering
    \includegraphics[width=\columnwidth]{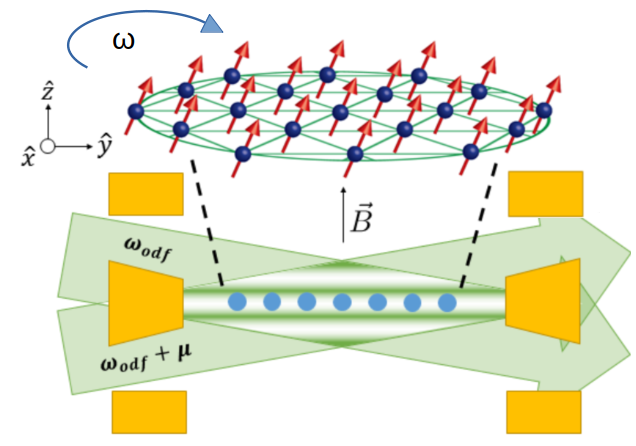}
    \caption{} 
    \label{fig:trap}
    \end{subfigure}
     \begin{subfigure}[t]{0.3\textwidth}
    \centering
    \includegraphics[width=\columnwidth]{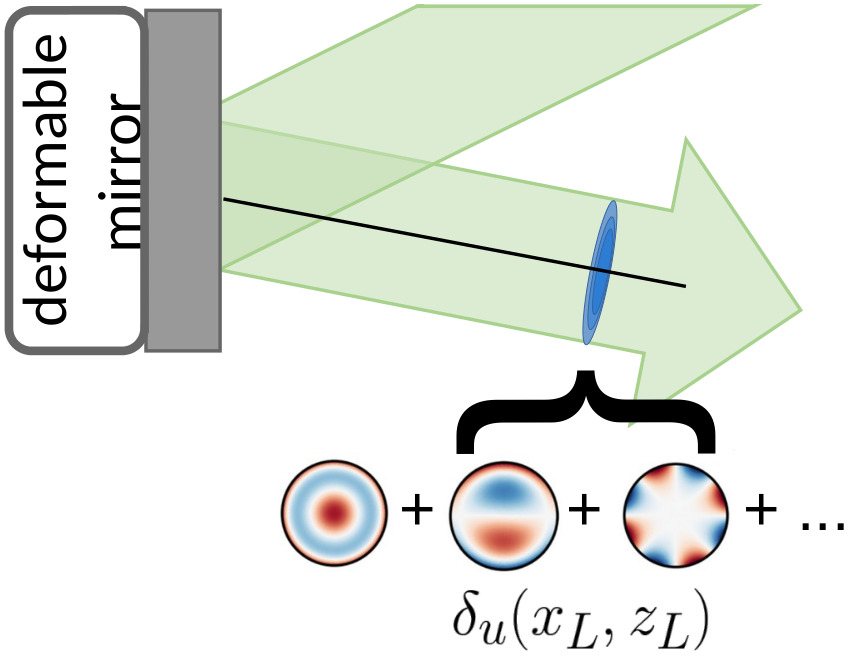}
    \caption{}  
    \label{fig:phase}
    \end{subfigure}
    \begin{subfigure}[t]{0.3\textwidth}
    \centering
    \includegraphics[width=\columnwidth]{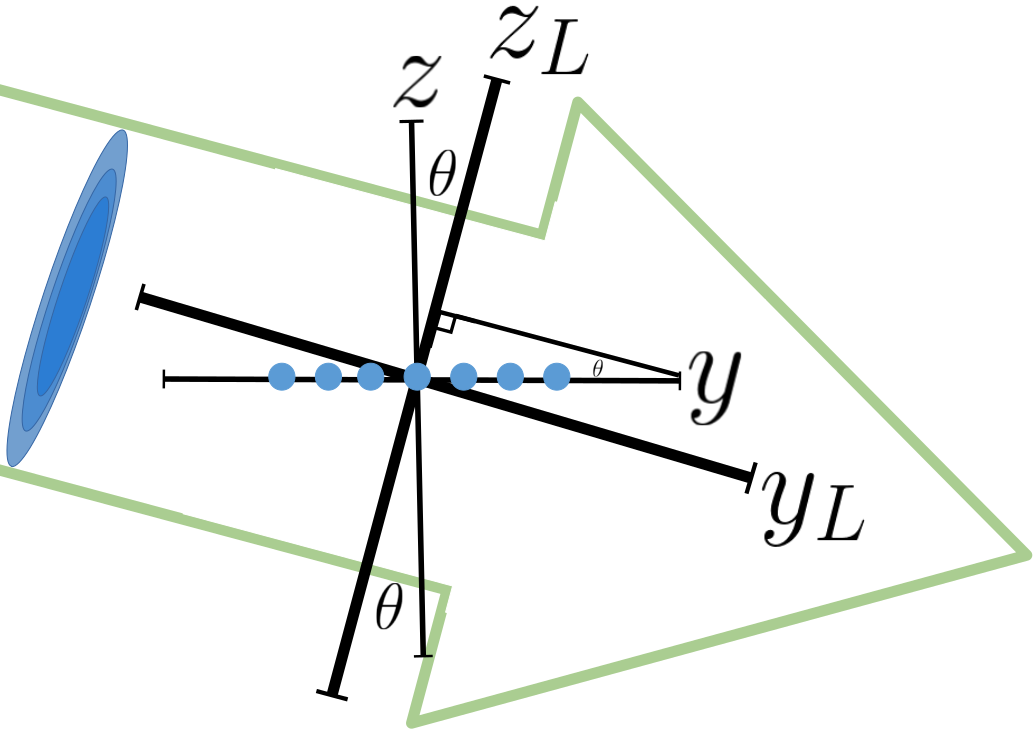}
    \caption{}  
    \label{fig:phase2}
    \end{subfigure}
    \captionsetup{justification=raggedright,singlelinecheck=false}
     \caption{ 
    (a) {\bf Ions in a Penning Trap.} The ions are axially confined by a set of cylindrical electrodes  and rotate with frequency $\omega$. In the presence of a strong magnetic field $\vec{B}$ along the $\hat{z}$ direction, the induced Lorentz force radially confines the ions. A moving 1D optical lattice  formed by interfering two beams with angular frequencies  $\omega_{\text{odf}}$ and $\omega_{\text{odf}} + \mu$ respectively, generates an optical dipole force. (b) {\bf Imprinting a Phase Pattern with a Deformable Mirror.} By reflecting one of the laser beams off of a deformable mirror (grey) surface, we can imprint a phase pattern on the wavefront (blue) of $\delta_{u}(x_L, z_L)$, where $x_L$, $y_L$ and $z_L$ are the beam-centric coordinates. The phase pattern $\delta_{u}(x_L, z_L)$ can be decomposed into a basis of Zernike polynomials. (c) To generate a distortion $\delta (x,y)$ at the ion crystal, one must generate a distortion in the upper beam given by $\delta_{u}(x_L, z_L)\equiv \delta(x,z_L/\sin \theta)$ with $\theta$ the angle between the beam propagation direction and the $y-$axis. The ions (blue) are at $z=0$.}
  \label{fig:PENNING}
\end{figure*}

Ions confined in RF traps are one of the leading platforms for quantum information processing \cite{Kielpinski2002,Brown2016,Bermudez_2017, Bruzewicz2019}. They hold the record for the highest fidelity entangling gates \cite{Gaebler2016,Ballance2016,Srivinas2021, Sawyer2021}, and have exceptional coherence times \cite{Wang2021}. However, scaling to larger numbers of ion qubits, potentially solved by photonic interconnects between Paul traps \cite{Monroe_2014}, or by shuttling ions in the QCCD architecture \cite{Kielpinski2002}, remains a key challenge. Penning ion traps offer the possibility of quantum information processing with samples of as many as 500 trapped ions self-assembled in a large 2D Coulomb crystal. They use a  set of cylindrical electrodes and static voltages to generate axial confinement. The radial confinement is provided by the Lorentz force experienced by the ions as they undergo a controlled rotation about the trap symmetry axis in the presence of a strong axial magnetic field, typically generated by a superconducting magnet (see Fig.\ref{fig:trap}). By encoding a qubit in two internal levels of the ions, Penning traps have the potential to perform quantum information processing with hundreds of qubits. These qubits typically have transition frequencies from 10's to 100~GHz, making microwaves suitable for global addressing \cite{Biercuk_2009}.

Two-qubit gates for entangling the ions have been engineered via spin-dependent optical dipole forces \cite{britton2012engineered}. These forces are generated by interfering two lasers with a difference frequency (sometimes called the beatnote) adjusted to excite phonons in the crystal (see Fig.\ref{fig:trap}). Virtual excitation of the center-of-mass mode, for example, generates collective spin-spin interactions across the ion array. 

Single-site rotations along with global rotations and a global entangling operation form a universal set of operations for quantum computation---that is, every unitary operation acting on a quantum register can be implemented \cite{schindler2013quantum}. Thus, for general quantum information processing in a Penning trap, what remains to be implemented and demonstrated is the ability to perform individual-qubit rotations.

A well known technique for introducing single-site qubit rotations with 1D ion strings is through AC Stark shifts with off-resonant, focused laser beams. AC Stark shifts produce $\sigma^Z$ rotations, which can be turned into more general rotations through the application of global rotations.

The analogous implementation of variable AC Stark shifts in a Penning trap is feasible but requires introducing focused co-rotating laser beams. The fast rotation frequency used in typical experiments  \cite{Bohnet2016,Gaerttner2017,Ball_2019,Mavadia2013,McMahon_2022}, ranging from tens to a few hundred kHz in recent NIST experiments, 
makes this task challenging. Variable AC Stark shifts can also be implemented with spatially fixed, focused beams directed at the correct radius so that an ion experiences a time-varying AC Stark shift as it rotates through the off-resonance laser beam. Similarly a fixed, focused pair of laser beams in a stimulated Raman configuration can be used to generate a spin rotation as an ion qubit passes through the laser beam waists. However, such approaches would in general require some sequential addressing of the ions, which is inherently slower than parallel addressing.

Here we propose another path for introducing variable AC Stark shifts that are static in the rotating frame of the ion crystal by using the same optical dipole force that is used for implementing a global entangling operation \cite{Bohnet2016,britton2012engineered}. This can be done by introducing distortions (or, more precisely, spatially dependent phase offsets) to the wavefront of the optical dipole force while setting the beatnote frequency to be a multiple of the rotation frequency (see Fig.~\ref{fig:phase} and Fig.~\ref{fig:phase2}). As we will see, a nice feature of this technique is that the rotations of the ions' spins can be conducted in parallel.

Any wavefront distortion on the unit disk can be decomposed into the basis of Zernike polynomials \cite{born2013principles}. Such functions are generally expressible as

\begin{align}\label{eq:zernike}
    Z_n^m(\rho, \phi) =
    \begin{cases}
     R_n^{|m|}(\rho)\cos{(m\phi)}, & \text{for } m \geq 0,\\
      R_n^{|m|}(\rho)\sin{(m\phi)}, & \text{for } m < 0,\\
    \end{cases} . 
\end{align}
Here $n \geq |m|$, $R_n^{|m|}(\rho)$ are radial polynomials defined on the unit disk. For example, the first few Zernike polynomials are

\begin{eqnarray}
    Z_0^0(\rho, \phi) = 1\\
    Z_1^{-1}(\rho, \phi) = 2\rho\sin (\phi)\\
    Z_1^1(\rho, \phi) = 2\rho\cos (\phi).
\end{eqnarray}

Through the decomposition of wavefront distortions in the basis of Zernike polynomials, we motivate two protocols in Sec.~\ref{sec:setup} for imprinting an AC Stark shift pattern across the crystal of ions. A numerical simulation of the protocols is outlined in Sec.~\ref{sec:simulation}, followed by a discussion of three primary sources of error in Sec.~\ref{sec:error}. These errors are from ignoring off-resonant terms, considering a finite number of Zernike polynomials, and applying too large of distortions. However, in Sec.~\ref{sec:numerics} we show with numerics that the errors can be controlled to have maximum infidelities as small as $10^{-3}$. In particular, we demonstrate the faithful reconstruction of an annulus, an elliptical Gaussian, and a displaced Gaussian using parameters that are representative of typical Penning  trap conditions  \cite{britton2012engineered,Bohnet2016,Gaerttner2017}. The annulus and elliptical Gaussian patterns produce initial states that are interesting for quantum simulation.  The displaced Gaussian is chosen to rotate a single qubit in the crystal. The high-fidelity reconstructions suggest that our protocol provides a path forward for implementing high-fidelity single- and multi-site qubit rotations in a Penning trap.

\section{Experimental Set-up and Protocols}\label{sec:setup} In a Penning trap, a system of $N$ ions is confined axially by voltages applied to a set of cylindrically symmetric electrodes. Radial confinement is implemented by the Lorentz force experienced by the ions as they rotate around the trap symmetry axis in the presence of a static $B_0\hat{z}$ field (see Fig.\ref{fig:trap}). The ion crystal rotation frequency is precisely controlled by a rotating electric field \cite{Huang1998}. In work at NIST \cite{britton2012engineered,Bohnet2016,Gaerttner2017}, the $\prescript{2}{}{S}_{1/2}$ ground-state valence electron spin $\ket{\uparrow} \equiv \ket{m_{J}=+1/2} (\ket{\downarrow} \equiv \ket{m_{J}=-1/2})$ in each trapped ${^9}$Be$^+$ ion encodes a spin-$1/2$ degree of freedom, which can be globally controlled by external microwaves resonant with the $124~$GHz frequency splitting of the electronic spin states in the $B_0=4.5$ T magnetic field of the trap. 

In a frame rotating with the ion crystal, the motion of the ions can be decomposed into in-plane modes, which describe ion motion in the plane of the ion crystal, and axial or drumhead modes, which describe ion motion transverse to the plane. For performing quantum simulations \cite{britton2012engineered,Bohnet2016,Gaerttner2017}, the drumhead modes are coupled to the spin degree of freedom by a spin-dependent optical dipole force (ODF) produced by a pair of off-resonant laser beams far detuned from the nearest optical transitions. The beams generate a one-dimensional (1D) traveling-wave lattice potential at a frequency $\mu$ (see Fig.~\ref{fig:trap}). The system can be well-described by the Hamiltonian
\begin{align}\label{eq:HODF}
    \hat{H}_{\text{ODF}}/\hbar = U\sum_i \cos{(\delta k \hat{z}_i - \mu t + \psi)}\hat{\sigma}_i^Z.
\end{align}
Here $U$ is the zero-to-peak AC Stark shift, $\delta k$ the wave vector of the moving lattice  potential, and $\psi$ is an initial phase. 
For clarity, we will use capital letters to denote directions in spin space, e.g. $\hat{\sigma}^Z$, and lowercase letters to denote directions in real space, e.g. $\hat{z}$.

In what follows we show how we can modify the ODF interaction in Eq.~\ref{eq:HODF} to introduce an AC Stark shift pattern that is static in the rotating frame of the crystal through the introduction of wave front deformations. We assume Eq.~\ref{eq:HODF} does not excite spin-dependent motion, which is reasonable if $\mu$ is far off-resonant with any modes.

It is convenient to assume that without a deformable mirror (DM) the two ODF beams have flat wavefronts. Then position dependent phase offsets (or wavefront distortions) $\delta(x, y)$ can be implemented in Eq.~\ref{eq:HODF} with a single DM that is inserted into one of the ODF laser beam lines, say the upper beam. The DM generates a position-dependent phase offset $\delta_u(x_L, z_L)$ where $x_L, z_L$ are the distances from the center of the beam in a coordinate system perpendicular to the k-vector of the laser beam (beam-centric coordinates, see Figs.~\ref{fig:phase}, \ref{fig:phase2}). We can choose the $x_L$ coordinate in the upper beam-centric coordinate system to be the same as the lab frame $x$ coordinate ($x_L=x$) at the single-plane crystal. Because the ODF beam crosses the ion crystal with an angle $\theta$, an ion located at the lab frame coordinates $(x,y)$ samples the DM generated phase offset $\delta_u(x, y\sin{\theta})$. Therefore to generate a wavefront offset pattern $\delta(x,y)$ at the ion crystal requires generating a wavefront pattern in the upper beam given by $\delta_u(x_L, z_L) \equiv \delta(x_L, z_L/\sin{(\theta)})$. In general the DM will be located at some distance from the ion crystal and an optical imaging set-up is required to image the DM surface to the ion crystal with different demagnification ratios in the $z_L$ and $x_L$ directions. 


Suppose, as sketched in the previous paragraph, we introduce waveform deformations to the 1D optical-dipole lattice potential (Eq.~\ref{eq:HODF}) so that the interaction is described by
\begin{align}\label{eq:HODF_deformations}
    \hat{H}_{\text{ODF}}/\hbar = U\sum_i \cos{(\delta k \hat{z}_i - \mu t + \psi + \delta(x_i, y_i))}\hat{\sigma}_i^Z.
\end{align}
Here $x_i = \rho_i \cos{(\phi_i^{\text{lab}}(t))}$ and $y_i = \rho_i\sin{(\phi_i^{\text{lab}}(t))}$ are the lab frame coordinates of ion $i$ in the $z=0$ plane, and $\delta(x, y)$ is a local phase shift that describes the distortions of the optical dipole force wavefronts. The azimuthal angle in the lab frame is related to the azimuthal angle in the rotating frame by $\phi^{\text{lab}}_i = \phi_i - \omega t$ where $\phi_i$ is independent of time for a stable crystal and $\omega$ is the ion crystal rotation frequency. 

Our goal is to employ wave front deformations $\delta(x,y)$ described in Eq.~\ref{eq:HODF_deformations} to generate an effective Hamiltonian given by
\begin{equation}\label{eq:desired_pattern}
    \hat{H}_{EFF}/\hbar = U\sum_i F(\rho_i, \phi_i)\hat{\sigma}_i^Z.
\end{equation}
Here $F$ describes an AC Stark shift pattern that is static in the rotating frame of the crystal. To do this, we will consider two different protocols with slightly different choices of $\delta(x, y)$. Before introducing these choices, we motivate a convenient decomposition of $F(\rho, \phi)$ by expanding in terms of Zernike polynomials, as described in Eq.~\ref{eq:zernike}. This gives a decomposition of the form

\begin{gather}
    F(\rho, \phi) = \sum_{n=0}^{\infty} \sum_{m=-\infty}^{\infty} Ac_n^m Z_n^m(\rho, \phi)\nonumber\\ = \sum_{n,m=0}^{\infty}AR_n^{m}(\rho) \Big( c_n^m  \cos{(m\phi)} + c_n^{-m}  \sin{(m\phi)}\Big)\nonumber\\ =
    \sum_{m=0}^{\infty}AP^m(\rho)\cos{(m\phi)}+ AQ^m(\rho)\sin{(m\phi)}\label{eq:phase_rewrite},
\end{gather}
where we have included $A$ as an overall amplitude scaling of the pattern $F(\rho, \phi)$, and set ${\rm max}(F(\rho, \phi))/A = 1.0$, which provides a normalization condition for the $c_n^m$. The radial functions $P^m(\rho)$ and $Q^m(\rho)$ are weighted sums of the appropriate $R_n^{|m|}(\rho)$ for $n\geq |m|$. Specifically, in expanding $F(\rho, \phi)$ in this way, we have gathered all terms with the same $\cos{(m\phi)}$ or $\sin{(m\phi)}$ azimuthal
dependence. 

We explore using the deformable mirror to generate each term in the sum of Eq.~\ref{eq:phase_rewrite}. Because $F$ in Eq.~\ref{eq:desired_pattern} can in general have arbitrarily high spatial frequency components (and the protocols we consider necessarily generate AC Stark shifts up to some maximum order $m_{\rm max}$) truncation error will be introduced by considering a finite number of terms in Eq.~\ref{eq:phase_rewrite}. This error will be considered in Secs.~\ref{sec:error} and \ref{sec:numerics} where we carry out a numerical analysis. 

We first consider generating an AC Stark shift pattern proportional to the term
\begin{equation}\label{eq:term}
    AP^m(\rho)\cos{(m\phi)}
\end{equation}
in Eq.~\ref{eq:phase_rewrite}. Suppose the DM is set to generate a distortion in Eq.~\ref{eq:HODF_deformations} of the form 
\begin{align}
    \tilde{\delta}(\rho, \phi^{\text{lab}}) = \delta^m_e(\rho)\cos{(m\phi^{\rm lab})}.
\end{align}
We now set $\hat{z}_i = 0$ (the ions are located in the $z=0$ plane) and substitute $\phi^{\text{lab}}_i = \phi_i - \omega t$. Then Eq.~\ref{eq:HODF_deformations} can be written
\begin{align}\label{eq:HODF_exponential}
    \hat{H}_{\text{ODF}}/\hbar = \frac{U}{2}\sum_i\big\{e^{i[-\mu t + \psi + \tilde{\delta}(\rho, \phi_i - \omega t)]} + c.c\big\}\hat{\sigma}^Z_i.
\end{align}
The phase modulation term can be expanded in terms of Bessel functions, using the  Jacobi–Anger expansion ($e^{iz\cos{\theta}}=\sum _{n=-\infty}^{\infty}  i^n J_n(z) e^{in\theta}$),
\begin{widetext}
\begin{align}\label{eq:HODF_exponential_expanded}
    \hat{H}_{\text{ODF}}/\hbar = \frac{U}{2}\sum_i\big\{e^{i[-\mu t + \psi]}\sum_{n=-\infty}^{\infty} i^n J_n(\delta^m_e(\rho_i))e^{in(m\phi_i
    -m\omega t)} + c.c.\big\}\hat{\sigma}^Z_i.
\end{align}
\end{widetext}
By setting $\mu=m\omega$ only the $n=-1$ term is static in the rotating frame. All other terms are rapidly oscillating and can be ignored. It is also possible to get static terms by choosing $\mu$ to be a higher integer multiple of $m\omega$. However, these terms will be scaled by a higher order Bessel function, and therefore produce a smaller static AC Stark shift, assuming sufficiently small arguments $\delta^m_e(\rho_i)$. Furthermore, we will show in Sec.~\ref{sec:rwa} and Appendix A that the contribution of the fast rotating terms is exactly zero if we choose to apply $H_{\rm ODF}$ for a duration $T$ satisfying $\omega T = 2\pi r$, with $r$ an integer value.

After some algebra we obtain,
\begin{align}\label{eq:HODF_RWA}
    \hat{H}_{\text{ODF}}/\hbar \approx U\sum_iJ_1(\delta^m_e(\rho_i))\sin{(m\phi_i
    - \psi)}\hat{\sigma}_i^Z.
\end{align}
Therefore by choosing a distortion of the form 
\begin{align}\label{eq:precomp}
 \delta^m_e(\rho) = J^{-1}_1(\frac{A}{2}P^m(\rho)),
\end{align}
Eq.~\ref{eq:HODF_deformations}, under the approximations discussed above, reduces to
\begin{align}\label{eq:stark_in_rotating}
    \hat{H}_{\text{ODF}}/\hbar \approx \frac{U}{2}\sum_iAP^m(\rho_i)\sin{(m\phi_i
    - \psi)}\hat{\sigma}_i^Z.
\end{align} 
With $\psi=-\pi/2$ this is exactly the targeted AC Stark shift pattern of Eq.~\ref{eq:term} (up to a factor of 1/2 which we introduce for convenience as we will discuss later).

In Eq.~\ref{eq:precomp}, we denote the choice of the wave front pattern described by $J_1^{-1}(\frac{A}{2}P^m(\rho))$ rather than $\frac{A}{2}P^m(\rho)$ as precompensation. Precompensation is always possible as long as $|\frac{A}{2}P^m(\rho)| \lesssim 0.58$, where 0.58 is the approximate maximum value that the $J_1$ Bessel function can take. This condition is always possible to satisfy by choosing $A$ small enough in Eq.~\ref{eq:precomp}. Note that this limits the phase offset $\delta^m_e(\rho)$ to be $|\delta^m_e(\rho)| \leq 1.84$. 

An identical derivation for odd Zernike polynomials reveals that a deformation of the form,
\begin{align}\label{eq:precomp2}
\delta(\rho, \phi^{\text{lab}}) = J_1^{-1}(\frac{A}{2}Q^m(\rho))\sin{(m\phi^{\text{lab}})}
\end{align}
gives,
\begin{align}\label{eq:HODF_RWA_odd}
    \hat{H}_{\text{ODF}}/\hbar \approx -\dfrac{U}{2}\sum_iAQ^m(\rho_i)\cos{(m\phi_i - \psi)}\hat{\sigma}_i^Z.
\end{align}
Additionally, for $m=0$ corresponding to a circularly symmetric pattern, Eq.~\ref{eq:HODF_exponential} reduces to 
\begin{equation}\label{eq:m0}
    \hat{H}_{\text{ODF}}/\hbar = U\sum_i \cos{(\delta^0_e(\rho_i) + \psi)}\hat{\sigma}_i^Z,
\end{equation}
so that choosing $\delta_e^0(\rho) = \cos^{-1}(AP^0(\rho)) - \psi$ for the $m=0$ terms reproduces the desired phase pattern Eq. \ref{eq:term}.

By sequentially setting the DM to generate each even ($AP^m(\rho) \cos{(m\phi)}$) and odd ($AQ^m(\rho) \sin{(m\phi)}$) term in Eq.~\ref{eq:phase_rewrite} for each $m$, the above derivation shows that one can apply any AC Stark shift pattern $F(\rho,\phi)$ (see Eq. \ref{eq:phase_rewrite}). However, sequential application can take a long time if there are many terms and the reset time of the DM is slow. Thus, it would be good to have a technique for applying all azimuthal phase patterns in parallel.

Applying in parallel means applying even and odd orders at the same time and applying different beat note frequencies at the same time. First, we show that we can apply both even and odd terms simultaneously, by considering a distortion of the form
\begin{equation}
    \delta(\rho, \phi^{\text{lab}}) = \delta^m_e(\rho)\cos{(m\phi^{\rm lab})} + \delta^m_o(\rho)\sin{(m\phi^{\rm lab})}\:.
\end{equation}
Using the Jacobi-Anger expansion, and setting $\mu = m\omega$, we find

\begin{widetext}
\begin{equation}
\label{eq:original}
    \hat{H}_{\text{ODF}}/\hbar =\frac{U}{2}\sum_i\Big(e^{i(-\mu t + \psi)}\exp{(i[\delta^m_e(\rho_i)\cos(m\phi_i - m\omega t) + \delta^m_o(\rho_i)\sin(m\phi_i - m\omega t)])} + c.c.\Big)\hat{\sigma}^Z_i,
\end{equation}
\begin{equation}\label{eq:doublesum}
    \hat{H}_{\text{ODF}}/\hbar =\frac{U}{2}\sum_i\Big(e^{i(-\mu t + \psi)}\sum_{a = -\infty}^\infty\sum_{b = -\infty}^\infty i^aJ_a(\delta^m_e(\rho_i))J_b(\delta^m_o(\rho_i))e^{i((a+b)(m\phi_i - m\omega t))}+ c.c.\Big)\hat{\sigma}^Z_i,
\end{equation}
Neglecting the fast rotating terms or operating with application times $T$ where $\omega T=2\pi r$ with $r$ an integer value (a condition at which the contribution of all non-static terms vanish), we obtain

\begin{equation}
\label{eq:bessel-products}
    \hat{H}_{\text{ODF}}/\hbar \approx U\sum_i\Big(\sum_{\substack{a + b = -1\\a,b\in\mathbb{Z}}}J_a(\delta^m_e(\rho_i))J_b(\delta^m_o(\rho_i))\cos{(a\frac{\pi}{2} - m\phi_i + \psi)}\Big)\hat{\sigma}^Z_i.
\end{equation}

For $a = -1$, and $\delta^m_o(\rho)=0$, this agrees with the expression in Eq.~\ref{eq:HODF_RWA}  ,
\begin{eqnarray}
\begin{split}
    \hat{H}_{\text{ODF}}/\hbar &\approx U\sum_iJ_{-1}(\delta^m_e(\rho_i))\cos{(-\frac{\pi}{2} - m\phi_i + \psi)}\hat{\sigma}^Z_i\\
    &=U\sum_iJ_{1}(\delta^m_e(\rho_i))\sin{(m\phi_i - \psi)}\hat{\sigma}^Z_i.
\end{split}
\end{eqnarray}

However, for both $\delta^m_e(\rho)$ and $\delta^m_o(\rho)$ non-zero, there are now terms given by higher-order Bessel functions that are static and non-zero. 

\begin{eqnarray}
   \hat{H}_{\text{ODF}}/\hbar &\approx& U\sum_i \Big (J_1(\delta^m_e(\rho_i))J_0(\delta^m_o(\rho_i))\sin({m\phi_i-\psi}) - J_1(\delta^m_o(\rho_i))J_0(\delta^m_e(\rho_i))\cos{(m\phi_i-\psi)} +\\&& J_1(\delta^m_e(\rho_i))J_2(\delta^m_o(\rho_i))\sin{(m\phi_i-\psi)} - J_2(\delta^m_e(\rho_i))J_1(\delta^m_o(\rho_i))\cos{(m\phi_i-\psi)} - ...\Big )\hat{\sigma}^Z_i.\nonumber
\end{eqnarray}

Fortunately, for small arguments, $J_n(x)\approx\frac{1}{n!}(\frac{x}{2})^n$ and $J_0(x)\approx 1-(\frac{x}{2})^2$, so that if we can choose $\delta^m_e(\rho_i)=A P^m(\rho_i)$ and $\delta^m_o(\rho_i)=A Q^m(\rho_i)$ to be small (by choosing $A$ to be small), the first two terms will reduce to the desired results (Eq.~\ref{eq:stark_in_rotating} and Eq.~\ref{eq:HODF_RWA_odd}). Note that we introduced the factor of 1/2 in the precompensation step (Eqs.~\ref{eq:precomp} and \ref{eq:precomp2}) to make the outcome of the serial and parallel protocols the same. The remaining terms will also be made small.
Explicitly:

\begin{eqnarray}
\label{eq:serial_expansion}
   \hat{H}_{\text{ODF}}/\hbar &\approx& \frac{U}{2}\sum_i\Big (AP^m(\rho_i)\sin({m\phi_i-\psi}) - AQ^m(\rho_i)\cos{(m\phi_i-\psi)}) + O(A^2)\Big) \hat{\sigma}^Z_i.
\end{eqnarray}

To first order in the arguments of the Bessel function, we see the even ($AP^m(\rho)\cos{ (m\phi)}$) and odd ($AQ^m(\rho) \sin{(m\phi)}$) terms in Eq.~\ref{eq:phase_rewrite} can be treated additively, and the Hamiltonian considered in Eq.~\ref{eq:original} can be used to apply both the even and odd  $m$ components in parallel. 

A similar analysis can be applied to show that it is possible to apply all different orders $m$ at the same time. For instance, consider the simplified case of two different nonzero even orders, $m_1$ and $m_2$, and suppose we set the beatnote frequency to $\mu = m_1\omega$. Then our wavefront deformation is given by 
\begin{equation}
     \delta(\rho, \phi^{\rm lab}) = AP^{m_1}(\rho)\cos{(m_1\phi^{\rm lab})} + AP^{m_2}(\rho)\cos{(m_2\phi^{\rm lab})}\label{comb}
\end{equation}
and our Hamiltonian is

\begin{equation}
    \hat{H}_{\text{ODF}}/\hbar =\frac{U}{2}\sum_i\Big(e^{i(-\mu t + \psi)}\exp{(i[AP^{m_1}(\rho_i)\cos(m_1\phi_i - m_1\omega t) + AP^{m_2}(\rho_i)\cos(m_2\phi_i - m_2\omega t)}]) + c.c.\Big)\hat{\sigma}^Z_i,
\end{equation}
\begin{equation}
    \hat{H}_{\text{ODF}}/\hbar =\frac{U}{2}\sum_i\Big(e^{i(-\mu t + \psi)}\sum_{a = -\infty}^\infty\sum_{b = -\infty}^\infty i^{a+b}J_a(AP^{m_1}(\rho_i))J_b(AP^{m_2}(\rho_i))e^{i(a(m_1\phi_i - m_1\omega t) + b(m_2\phi_i - m_2\omega t))}+ c.c.\Big)\hat{\sigma}^Z_i
\end{equation}
As before, neglecting the fast rotating terms, or operating with application times $T$ where $\omega T$ is a positive integer multiple of $2\pi$ (a condition at which all non-static terms vanish) we get 
\begin{equation}
    \hat{H}_{\text{ODF}}/\hbar \approx U\sum_i\Big(\sum_{\substack{am_1 + bm_2 = -m_1\\a,b\in\mathbb{Z}}}J_a(AP^{m_1}(\rho_i))J_b(AP^{m_2}(\rho_i))\cos{((a+b)\frac{\pi}{2} - m_1\phi_i + \psi)}\Big)\hat{\sigma}^Z_i.
\end{equation}

The lowest order terms occur when $a=-1$ and $b=0$, resulting in

\begin{equation}\label{eq:parallel_expansion}
    \hat{H}_{\text{ODF}}/\hbar \approx \frac{U}{2}\sum_i\Big(AP^{m_1}(\rho_i)\sin{(m_1\phi_i - \psi)} + O(A^2) ...\Big)\hat{\sigma}^Z_i. 
\end{equation}
\end{widetext}
For small $A$ this is approximately the desired AC Stark shift pattern. If we set the ODF beatnote $\mu = m_2\omega$ we select an AC Stark shift pattern described by the second term in Eq.~\ref{comb}.

When considering terms with $m=0$ and setting $\psi=-\pi/2$, the leading order contribution is instead $UAP^0(\rho_i)$, so that there is an additional factor of $2$ multiplying the radial polynomial. Note that the precompensation for $m=0$ (see Eq.~\ref{eq:m0}) was chosen to also make the outcome of the serial and parallel protocols the same.

The above analyses support two experimental procedures for generating an AC Stark shift pattern $F(\rho, \phi)$ that is static in the rotating frame of the crystal. The first is sequential: for a phase pattern with terms of at most order $m_{\rm max}$, we sequentially set the DM to $2m_{\rm max} + 1$ different azimuthal phase patterns, applying the appropriate beatnote frequency at each step and the corresponding precompensation in the applied waveform. This has the advantage of allowing for larger amplitudes $A$ and higher accuracy.

The second procedure is a  parallel application: we set the DM once to a phase pattern proportional to $F(\rho, \phi)$ at the ion crystal, and simultaneously (or in rapid succession) apply all beatnote frequencies $\mu_m=m\omega$, for $0 \leq m \leq m_{\rm max}$. The beatnote at $\mu_m$ will imprint an AC Stark shift in the rotating frame of the ions proportional to $P^m(\rho)\cos (m\phi) + Q^m(\rho)\sin (m\phi)$, rotating the ion's spins according to the $m^{\rm th}$-order component of $F(\rho, \phi)$. This has the benefit of being faster if the pattern has a large number of frequency components, but at the cost of lower accuracy and requiring smaller amplitudes. We note that small amplitudes (i.e. $A$ in Eq.~\ref{eq:phase_rewrite} or $\delta$ in Eq.~\ref{eq:HODF_deformations}) can be offset through the use of large $U$ or long application times $T$.

\section{Numerical Simulation}\label{sec:simulation}
In this section we outline a numerical study whose results are presented in Secs.~\ref{sec:error} and \ref{sec:numerics} for preparing arbitrary qubit rotation profiles, $F(\rho, \phi)$, across the crystal. In the work that follows, we set $U=2\pi\times 10$~kHz and $\omega=2\pi\times180$~kHz, which are typical experimental parameters \cite{Bohnet2016, Gaerttner2017}. Our goal will be to prepare the ions in the state 
\begin{equation}\label{eq:crystal_state}
    \ket{\psi(T)} =  \bigotimes_i e^{-i UF(\rho_i, \phi_i)T\hat{\sigma}_Z^i}\ket{+}_i,
\end{equation}
where $T$ is the gate time.

First, we will prepare all of the ions in the $\ket{+} = \frac{1}{\sqrt{2}}(\ket{\uparrow} + \ket{\downarrow})$ state, which can be easily done by preparing all ions in $\ket{\downarrow}$ and then applying a global rotation around the $Y$ axis. We will then determine a maximum $m$ and $n$, based on the desired fidelity of the state preparation, such that we approximately reconstruct $F$ as $\tilde{F}$, using only $Z_{m}^{n}$ for all $|m| \leq m_{\rm max}$ and $n \leq n_{\rm max}$. Writing $\alpha_n^m$ for the coefficients of $F$ in the Zernike basis, we have
\begin{equation}\label{eq:expansion}
    F = A \sum_{\substack{-n \leq m \leq n\\0\leq n \leq \infty}}\alpha_n^mZ_n^m \approx A \sum_{\substack{-m_{\rm max} \leq m \leq m_{\rm max}\\ 0\leq n \leq n_{\rm max}}}\alpha_n^mZ_n^m =    \tilde{F}.
\end{equation}
\newline
Experimentally, $n_{\rm max}$ could be constrained by the available resolution of the deformable mirror---since an $n^{th}$ order polynomial is determined by $n + 1$ points, a mirror with $N$ actuators in a dimension can only hope to parameterize a family of polynomials of degree $N - 1$. In our analysis we will assume that the DM has a sufficiently large number of actuators and prioritize minimizing $m_{\rm max}$, which sets the number of terms in the decomposition of $F$ (see Eq.~\ref{eq:phase_rewrite}) that will be included in the reconstruction $\tilde{F}$. Larger $m_{\rm max}$ in general requires a longer gate time or higher laser power.

As discussed in Sec.~\ref{sec:setup}, there are two ways to apply the full phase pattern---sequentially, and in parallel. In the analysis that follows we will consider both of these approaches. Although in principle one could consider using these techniques to apply arbitrary qubit rotations, for the purposes of assessing the performance of our protocols we consider the experimentally useful example of $\pi$ rotations. That is, in the examples considered in  Secs.~\ref{sec:error} and \ref{sec:numerics} the time evolution will be set for a time $T$ such that in the final state $\ket{\psi(T)}$ the ion located at the maximum of the phase-pattern $F$ will be rotated by $\pi$ radians in the $XY$-plane.

\section{Sources of Error}\label{sec:error}
We will now discuss three sources of error that can occur in the protocols described in Sec.~\ref{sec:setup}. To quantify the error and to analyze the performance of our protocol we will use the single-spin infidelity 
\begin{equation}\label{eq:infidelity}
    I_j = 1 - |\braket{\mathbb{P}^j\tilde{\psi}(T)|\mathbb{P}^j\psi(T)}|^2.
\end{equation}
where $\mathbb{P}^j$ traces out all but the $j^{th}$ ion. The quantity $I_j$ is the infidelity of spin j in the simulated state  $\ket{\tilde{\psi}(T)}$ with respect to the target state $\ket{\psi(T)}$.  State-of-the-art quantum information processing platforms often have single-qubit gate infidelities of $10^{-3} - 10^{-2}$\cite{Ballance2016, Gaebler2016, hong2020demonstration, arute2019quantum} or less, and so this will be the standard of comparison in our analysis.

\subsection{Rotating Wave Approximation}\label{sec:rwa}
\begin{figure*}
\begin{subfigure}[t]{0.45\textwidth}
    \centering
  \includegraphics[width=\columnwidth]{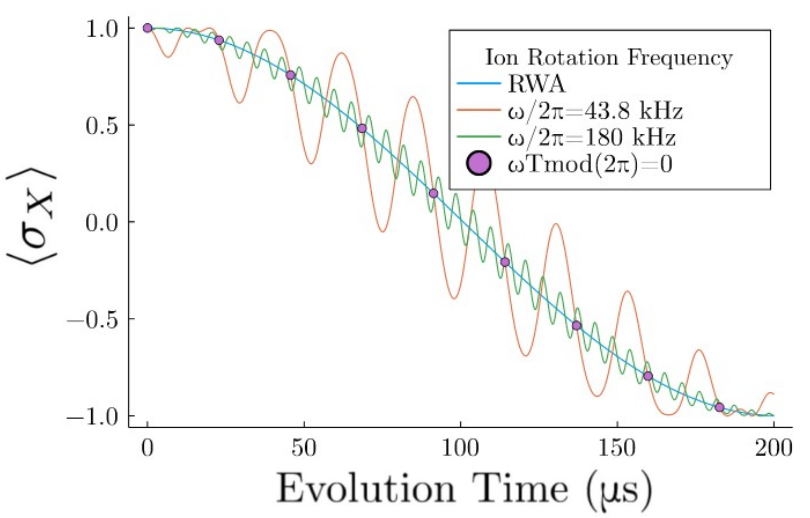}
  \caption{}
  \label{fig:rwa_oscillations}
  \end{subfigure}
\begin{subfigure}[t]{0.45\textwidth}
    \centering
  \includegraphics[width=\columnwidth]{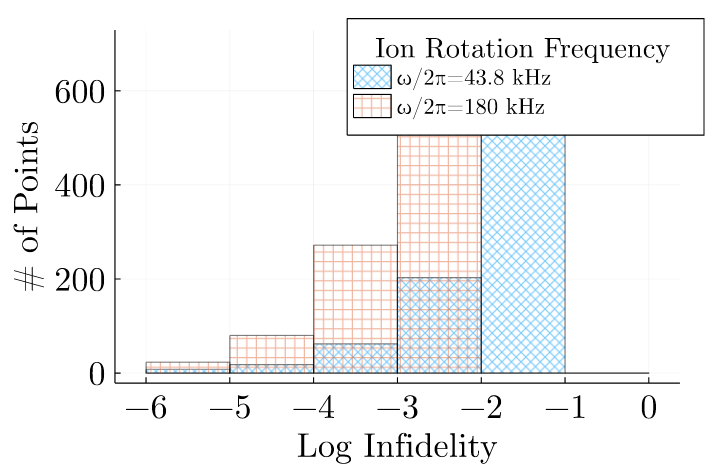}
    \caption{}
  \label{fig:histogram}
  \end{subfigure}
  \captionsetup{justification=raggedright,singlelinecheck=false}
  \caption{(a) {\bf Error from the Rotating Wave Approximation.} Starting in the state $\ket{\psi(0)} = \frac{1}{\sqrt{2}}(\ket{\uparrow} + \ket{\downarrow})$, we plot $\braket{\sigma_X}$ as a function of time under evolution by Eq.~\ref{Eq} with $\mu=\omega$ and $\delta(\rho, \phi) = A\rho\cos{(\phi)}$ for $A=0.25$,  and  different values of $\omega$ at $(\rho,\phi)=(1, 0)$. These expectation values are compared with those coming from the rotating wave approximation given by Eq.~\ref{eq:HODF_RWA} with $m=1$. We see that by increasing the ion rotation frequency from $\omega = 2\pi \times 43.8$~kHz to $\omega = 2\pi \times 180$~kHz, the  correction terms in Eq.~\ref{eq:rwa_time} get  suppressed so that the maximum infidelity is no larger that $10^{-2}$. (See Fig.~\ref{fig:histogram}). The pink dots mark times that are positive integer multiples of $2\pi$/$43.8$kHz where the RWA error vanishes for the slower trap rotation frequency.  (b) {\bf Distribution of Errors from RWA.} Figure~\ref{fig:rwa_oscillations} samples $\braket{\sigma_X}$ at 1000 different points in time. Figure~\ref{fig:histogram} shows the histogram of the log infidelities between the exact evolution and evolution under Eq.~\ref{eq:HODF_RWA} evaluated at the same points in time. Choosing $\omega=2\pi\times 43.8$~kHz results in most of the infidelities being larger than $10^{-2}$. Increasing the rotation frequency to $\omega=2\pi\times 180$~kHz, the maximum infidelity is decreased to smaller than $3\times10^{-3}$(orange). By sampling many different values of evolution time, we can be confident that the infidelity contributed by the RWA for arbitrary angles of rotation are sufficiently small for high-fidelity generation of $\ket{\psi(T)}$~(Eq.~\ref{eq:crystal_state}).}
\label{fig:RWA_error}
\end{figure*}

The first type of error comes from the rotating wave approximation, where we ignore the rapidly oscillating terms in Eq.~\ref{eq:HODF_exponential_expanded}. As we will show  below these errors nevertheless can be avoided if one chooses the gate time to be commensurate with the ion crystal rotation frequency: $\omega T= 2\pi r$, with $r$ an integer. 

As a simple example, we consider a phase pattern with a single angular order $m$ and with an additional amplitude parameter $A$ that will be chosen sufficiently small so that precompensation is not necessary. Thus our phase function in Eq.~\ref{eq:HODF_deformations} is given as
\begin{equation}\label{eq:rwa_phase}
    \delta(\rho, \phi) = A P^m(\rho)\cos{(m\phi)}.
\end{equation}
For our simple Hamiltonian, we can analytically express the expectation value $\langle\sigma_X\rangle$ of the spin after evolving for time $T$. Writing $f(\tau)$ as the time-dependent coefficient in Eq.~\ref{eq:HODF_exponential_expanded}, i.e. $\hat{H}_{\text{ODF}}/\hbar = \sum_j f_j(t)\hat{\sigma}^Z_j$,
and setting $\mu = m\omega$, we find (for $m\neq0$)
\begin{widetext}

\begin{equation}
\braket{\tilde{\psi}(T)|\sigma_X^j|\tilde{\psi}(T)} = \cos{\Big (2\int_0^{T}f_j(\tau) d\tau\Big)}. \label{Eq}
\end{equation}
We can write
\begin{equation}
    \int_0^T d\tau f_j(\tau) =  \sum_{n=-\infty}^{\infty} s_j(n),
\end{equation}
where
\begin{equation}\label{eq:analytic}
    s_j(n) = 
    \begin{cases}
    U J_1(AP^m(\rho_j)) \sin[m\phi_j-\psi]T , & x\text{if } n = -1\\
  \frac{4 U J_n(AP^m(\rho_j)) \sin
   \left(\frac{1}{2} m (n+1) T \omega\right) \cos \left(m
   n \phi +\frac{1}{2} (\pi  n-m (n+1) T \omega)+\psi
   \right)}{m (n+1) \omega}   & \text{if }n \neq -1.\\
      \end{cases}
\end{equation}

Therefore, if $UJ_n(AP^m(\rho_j))/(m\omega) \ll 1$ then, 
\begin{align}\label{eq:rwa_time}
\braket{\tilde{\psi}(T)|\sigma_X^j|\tilde{\psi}(T)} &\approx \cos{(\Theta)} - \sin{(\Theta)}\sum_{n = -\infty, n \neq -1} ^\infty a_n(T (n+1) m\omega)\frac{U}{(n+1) m\omega}\\
\Theta &=  2 UJ_1(AP^m(\rho_j))   \sin[m\phi_j-\psi]T, 
\end{align}

\end{widetext}
for some function $a_n$. For $m=0$ we have simply that 
\begin{equation}
    \int_0^{T}f_j(\tau) d\tau = U\cos{(AP^0(\rho_j)+\psi)}T,
\end{equation}
which has no time-dependent corrections.

As an example, choosing $m=1$ and $P^1(\rho) = \rho$, we will evaluate Eq.~\ref{Eq} at the point that will have the largest infidelity, $\rho=1$ and $\phi=0$, with $A = 0.25$. This choice of $A$ allows us to focus primarily on effects from the RWA, and ignore the other sources of error discussed below. This gives superimposed oscillations around a cosinuisoidal evolution, with corrections proportional to $\frac{U}{m \omega}$. Thus, by increasing $\omega$ for fixed $U$, the RWA becomes more accurate. This is shown in Fig.~\ref{fig:rwa_oscillations}, where the evolution is sampled at 1000 points in time. As $\omega$ is increased from $2\pi\times43.8$~kHz (blue) to $2\pi\times180$~kHz (orange), the oscillations become smaller. We see  that the ion completes a full $\pi$ rotation in nearly 200~$\mu$s, as expected from Eq.~\ref{eq:HODF_RWA}. Additionally, we see that at evolution times that are positive integer multiples of $2\pi/\omega$ the difference between the exact evolution and the RWA is zero (see pink points in Fig.~\ref{fig:RWA_error}.). This can be seen in Eq.~\ref{eq:analytic}, as  the terms with $n\neq -1$ are zero at  these points. See also Appendix A for a more generic case.  In Sec.~\ref{sec:numerics} we choose evolution times that take advantage of this fact.
Note that for this simple case the fast-rotating terms also vanish at  $T=\pi/\omega$. 

Figure~\ref{fig:histogram} shows a histogram of the log infidelity obtained at the different evolution times sampled in Fig.~\ref{fig:rwa_oscillations}. By increasing the rotation frequency from $2\pi\times43.8$~kHz (blue) to $2\pi\times180$~kHz (orange), the maximum single-spin infidelity is decreased from approximately $5\times10^{-2}$ to $3\times10^{-3}$. This implies that setting $\omega=2\pi \times 180$~kHz is sufficient for obtaining infidelities of $3\times10^{-3}$ or smaller.

\subsection{Truncation}\label{sec:truncation}
The second source of error comes from the fact that in practice we apply a finite number of beatnote frequencies, truncating the basis expansion at finite order. This  will produce imperfect reconstructions of the desired phase pattern. While choosing finite $n_{\rm max}$ is also a possible source of error, as disussed earlier we choose $n_{\rm max}$ large enough that it is not the limiting factor---effectively assuming that the DM has enough actuators to give good resolution.

The Zernike polynomials form an orthogonal set of polynomials on the disk, $D$, and therefore arbitrary functions can be decomposed into these polynomials. 
The inner product on the space of functions on the disk is given by:
\begin{equation}
\langle F, G\rangle = \int_D \rho d\rho d\theta F\cdot G
\end{equation}
We can write the coefficients of $F$ from Sec.~\ref{sec:simulation} as
\begin{equation}\label{eq:exact_coefficients}
    \alpha_n^m = \frac{2n + 2}{\epsilon_m\pi}\langle \frac {F}{A}, Z_n^m\rangle,
\end{equation}
where $\epsilon_m$ is 2 if $m=0$, and 1 otherwise. The prefactor in Eq.~\ref{eq:exact_coefficients} is due to the fact that the polynomials are not normalized,
\begin{equation}
    \langle Z_n^m(\rho, \phi)Z_{n'}^{m'}(\rho, \phi)\rangle = \frac{\epsilon_m\pi}{2n + 2}\delta_{n,n'}\delta_{m, m'}.
\end{equation}

Truncating the number of terms we include will give us a different phase function, $\tilde{F}$, from which we can define the error from truncation as 
\begin{equation}
\mathcal{E} = \text{max}_D(|F - \tilde{F}|/A).
\end{equation}
 Note that max$(F(\rho,\phi))=A$ (see discussion after Eq.~\ref{eq:phase_rewrite}), so $\mathcal{E}$ is the truncation error normalized to the maximum value of $F$.

The amount that this truncation contributes to the infidelity will vary depending on the particular phase function $F$ being considered, as we will see in Sec.~\ref{sec:numerics}. Here, to provide a rough estimate, we consider the maximum single-spin infidelity $\varepsilon$ across the crystal, defined as
\begin{eqnarray}
    \varepsilon &=& \text{max}_j(I_j)\nonumber\\
   & =& \text{max}_j(1 - |\braket{\mathbb{P}^j\tilde{\psi}(T)|\mathbb{P}^j\psi(T)}|^2)\nonumber\\
   & \approx& \text{max}_j\Bigg(1 - \Big|1 - \frac{1}{2}\Big(UAT(F(\rho_j, \phi_j)-\tilde{F}(\rho_j, \phi_j))/A\Big)^2\Big|^2\Bigg)\nonumber\\
   & \approx& (\mathcal{E}UAT)^2,\label{trunc}
\end{eqnarray}
where $\mathbb{P}^j$ traces out all but the $j^{th}$ ion.

We consider rotations where the ion located at the maximum of $F$ is rotated by $\pi$ radians, corresponding to $UAT = \pi/2$. An infidelity requirement of $\varepsilon$ therefore in general necessitates a truncation error $\mathcal{E} \lessapprox \frac{2}{\pi}\sqrt{\varepsilon}$. For an infidelity requirement of $10^{-2}$ ($10^{-3}$) the maximum truncation error should be less than $0.064$ ($0.02$). 

\subsection{Linear Approximation}\label{sec:linear-approx}
The final source of infidelity, which is only relevant for the parallel application discussed in Sec.~\ref{sec:setup}, is in assuming  that $A$ in Eq.~\ref{eq:serial_expansion} and Eq.~\ref{eq:parallel_expansion} is small enough so  that ignoring higher order terms is justified. By increasing the product $UT$ and decreasing the amplitude $A$, the linear approximation can be made arbitrarily good. However increasing $UT$ will also increase decoherence due to off-resonant light scattering from the ODF beams \cite{Uys2010} during the qubit rotations.

To estimate the contribution of higher order terms in $A$ on the infidelity for the target case of a $\pi$-rotation on a single ion, we consider the parallel application
of two $P^{m_1},P^{m_2}$ terms (see Eq.~\ref{comb}). It can be shown that leading corrections, of order $O(A^2)$, arise when $2m_1=m_2$. In this case the leading order static terms generate a Hamiltonain of the form

\begin{widetext}
\begin{equation}
    \hat{H}_{\rm ODF}/\hbar \approx \frac{U}{2}\Big (\sum_i AP^{m_1}(\rho_i)\sin{(m_1\phi_i - \psi)}-\frac{A^2}{2} P^{m_1}(\rho_i)P^{2 m_1}(\rho_i))\cos{(m_1\phi_i - \psi)})+ O(A^3)\Big)
\end{equation}
\end{widetext}

We now compute the infidelity between the state $\ket{\phi} = e^{-i\hat{H}_{\rm ODF}T/\hbar}\ket{+}$ and $\ket{\tilde{\phi}} = e^{-i \frac{UAT  P^{m_1}(\rho_0)}{2}\hat{\sigma}_Z}\ket{+}$ obtained by considering only the desired first order term. We choose $\psi=-\pi/2$ in the rest of this discussion. By Taylor expanding, we find   the infidelity to be

\begin{eqnarray}
    I(\tilde{\phi}, \phi) &\approx& 
\Big|\frac{ U T }{4}A^2 P^{m_1}(\rho_i)P^{2m_1}(\rho_i)\sin{(m_1\phi_i)} \Big|^2
\end{eqnarray}

Thus, for  $P^{m_1}(\rho_i)P^{2m_1}(\rho_i)\sin{(m_1\phi_i)}<1$, and $UT A/2\sim \pi$, the infidelity reduces to

\begin{eqnarray}
    I(\tilde{\phi}, \phi) &\lesssim& 
(\frac{ \pi}{2}A)^2.
\end{eqnarray} 
We see that if $A \leq 0.02$ the infidelity can be constrained to be less than $10^{-3}$. If we relax our infidelity requirements to $10^{-2}$ we can choose $A$ as large as $A=0.06$. While these amplitude requirements may seem strict, we note that this estimate is pessimistic - we have considered the worst-case situation when $2m_1 = m_2$, which gives leading order error contributions of size $O(A^2)$. As we will see in Sec.~\ref{sec:numerics}, the amplitude can often be made larger. In fact, as we will see in Sec.~\ref{sec:edisplaced} where we look at the case of flipping the spin of a single ion, the amplitude can be taken more than an order of magnitude larger while achieving the same infidelity goals.

\section{Numerical Results}\label{sec:numerics}
 Following the discussion of Secs.~\ref{sec:simulation} and \ref{sec:error}, we now numerically demonstrate a few interesting examples of implementing different AC Stark shift patterns across a circular crystal with our protocols from Sec.~\ref{sec:setup}.
 Preparation of initial states with targeted spatial profiles can be of great utility for investigating propagation of quantum information and entanglement. With that purpose in mind, here we consider a range of geometries including an annulus, an elliptical Gaussian, and a displaced Gaussian. For these patterns results for $\langle\sigma_X\rangle$ as well as the log infidelity across a crystal of $91$ ions are shown. This number was chosen to have inter-ion spacings of $0.1$ of the crystal diameter. For the phase patterns considered, Fig.~\ref{fig:annulus_error}, Fig.~\ref{fig:elliptical_error} and Fig.~\ref{fig:displaced_error} show the truncation error, $|\mathcal{E}|$, from considering a finite number of Zernike polynomials. Next, we study the error generated during the dynamical evolution. 
First we apply the protocol in series, evolving under each even ($AP^m(\rho) \cos{ (m\phi)}$) and odd ($AQ^m(\rho) \sin{(m\phi)}$) term in Eq.~\ref{eq:phase_rewrite} that constitutes $\tilde{F}(\rho, \phi)$ one-by-one, for two different choices of target maximum infidelity. The infidelities of the final state are shown in Figs.~\ref{fig:annulus_infid},~\ref{fig:elliptical_infidelity}, and~\ref{fig:displaced_gaussian}. We discuss contributions to the infidelity arising from the RWA and truncation errors (see Sec.~\ref{sec:error}). Next, we apply the protocol in parallel for all $m$ such that $0 \leq m \leq m_{\rm max}$ in Eq.~\ref{eq:expansion}. In  Fig.~\ref{fig:pelliptical} and Fig.~\ref{fig:pdisplaced} we show the corresponding infidelities for the elliptical and displaced Gaussians. (The annulus only requires implementing a single $m=0$ term.)

\subsection{Annulus}\label{sec:ann_sec}
As a first example, we consider preparing ions in an annulus. To make the problem of reconstruction in a basis of continuous functions easier, we will smooth the edges with sigmoid functions, giving
\begin{equation}
    g(\rho) = \frac{1}{1 + e^{-\kappa(\rho-r_1)}} -\frac{1}{1 + e^{-\kappa(\rho-r_2)}}.
\end{equation}
Scaling this function to be one at its maximum, we have a targeted normalized AC Stark shift pattern given by:
\begin{equation}
F(\rho, \phi) = Ag(\rho)/g(r_1 + \frac{1}{2}(r_2-r_1)),
\end{equation}
which corresponds to the phase function 
\begin{equation} \label{eq:annulus_pre}
    \delta(\rho, \phi) = \cos^{-1}{(F(\rho, \phi))}
\end{equation}
in Eq.~\ref{eq:HODF_deformations}. For our numerical experiment we will set $r_1=0.45$, $r_2=0.55$, and $\kappa=10$. This value of $\kappa$ was chosen to avoid sharp rising and falling edges for the annulus.

\subsubsection{Reconstruction}\label{sec:annulus_reconstruction}
\begin{figure}
    \centering
    \includegraphics[scale=.25]{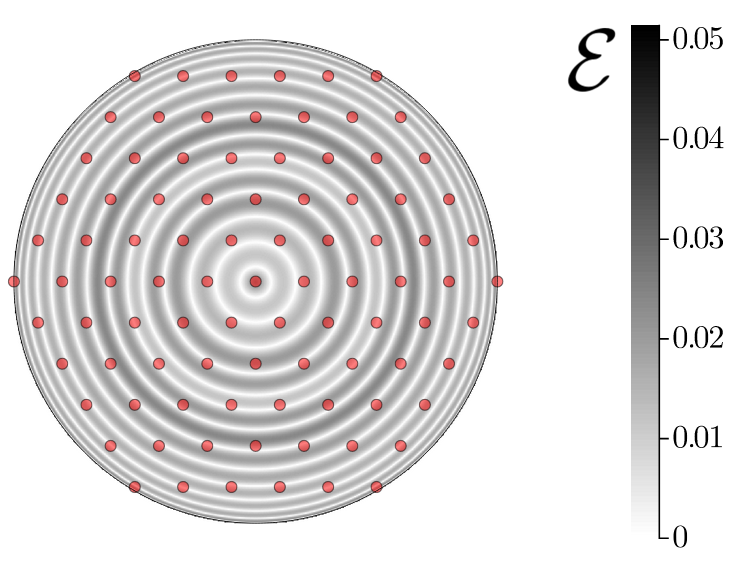}
    \captionsetup{justification=raggedright,singlelinecheck=false}
    \caption{Absolute value of the error $\mathcal{E}=|F - \tilde{F}|/A$ in reconstructing an annulus. With $n_{\rm max}= 24$ and $m=0$ we can reconstruct an annulus with an error no larger than $0.05$. This is sufficiently low error to reproduce a high fidelity state. (See Fig.~\ref{fig:annulus_infid}.) The red dots represent the ion positions and are shown for reference.}
    \label{fig:annulus_error}
\end{figure}

Because the phase pattern is azimuthally symmetric, the only nonzero coefficients have $m=0$ and only the application of a single beatnote with frequency $\mu=m\omega=0$ is required. We see that all terms are static, and therefore incur no error from the RWA. We choose $n_{\rm max}$ to be sufficiently large ($n_{\rm max} = 24$) so that the reconstruction error $\mathcal{E}$, presented in Fig.~\ref{fig:annulus_error}, is less than 0.064. From the discussion in Sec.~\ref{sec:error} this should enable a single-spin infidelity of less than $10^{-2}$.    
The next contributing error term is radially symmetric, since the pattern itself is radially symmetric, which can be seen clearly in Fig.~\ref{fig:annulus_error}. 

\subsubsection{Evolution}\label{sec:annulus_parallel}
Given that all the terms in the Zernike expansion have the same value of $m$, they can be applied simultaneously using the precompensation technique (Eq.~\ref{eq:annulus_pre}) above,  incurring no errors from the linear approximation. Setting $A=1.0$, we obtain a  gate time of $25~\mu$s for $U=2\pi\times10$~kHz, which is significantly faster than typical decoherence times in trapped ions.

\begin{figure}
    \centering
    \includegraphics[scale=.25]{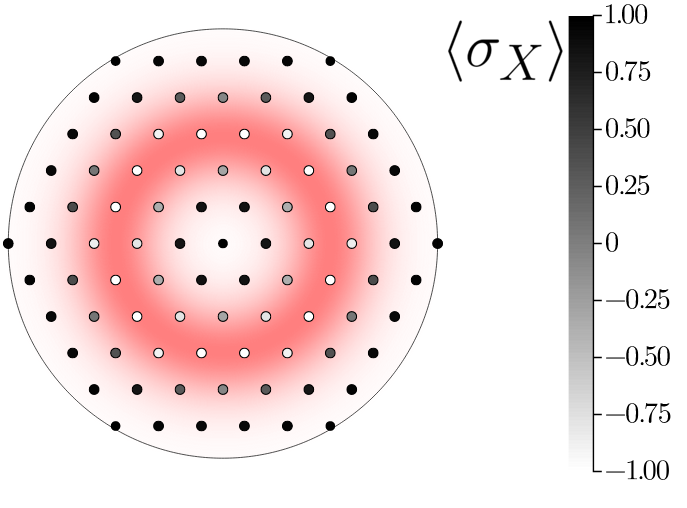}
    \captionsetup{justification=raggedright,singlelinecheck=false}
    \caption{$\langle\sigma_X\rangle $  for an annulus after following the protocol in Sec.~\ref{sec:simulation}. Using $n_{\rm max}=24$ and $m=0$, we see that $\langle\sigma_X\rangle $ is $-1$ on the annulus and $1$ outside the annulus, as desired.The pink pattern illustrates the targeted AC Stark shift pattern $F$. }
    \label{fig:annulus_expectation}
\end{figure}

\begin{figure*}
  \begin{subfigure}[t]{0.3\textwidth}
    \centering
    \includegraphics[scale=.2]{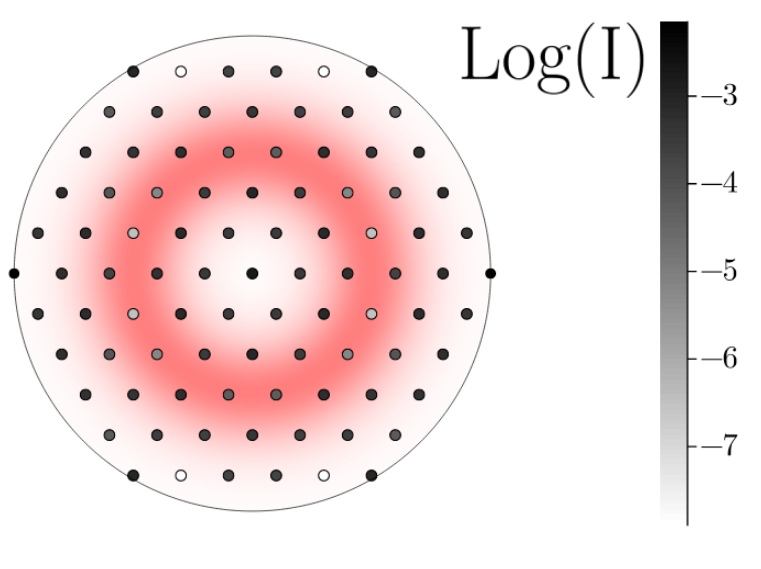}
    \caption{\textbf{}}
    \label{fig:annulus_infid1}
    \end{subfigure}
    \begin{subfigure}[t]{0.3\textwidth}
    \centering
    \includegraphics[scale=.2]{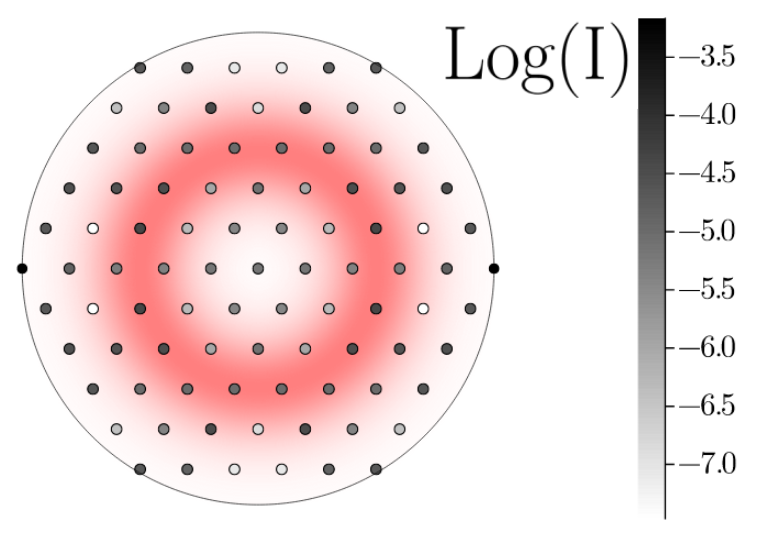}
    \caption{\textbf{}}
    \label{fig:annulus_infid2}
    \end{subfigure}
    \begin{subfigure}[t]{0.3\textwidth}
    \centering
    \includegraphics[scale=.14]{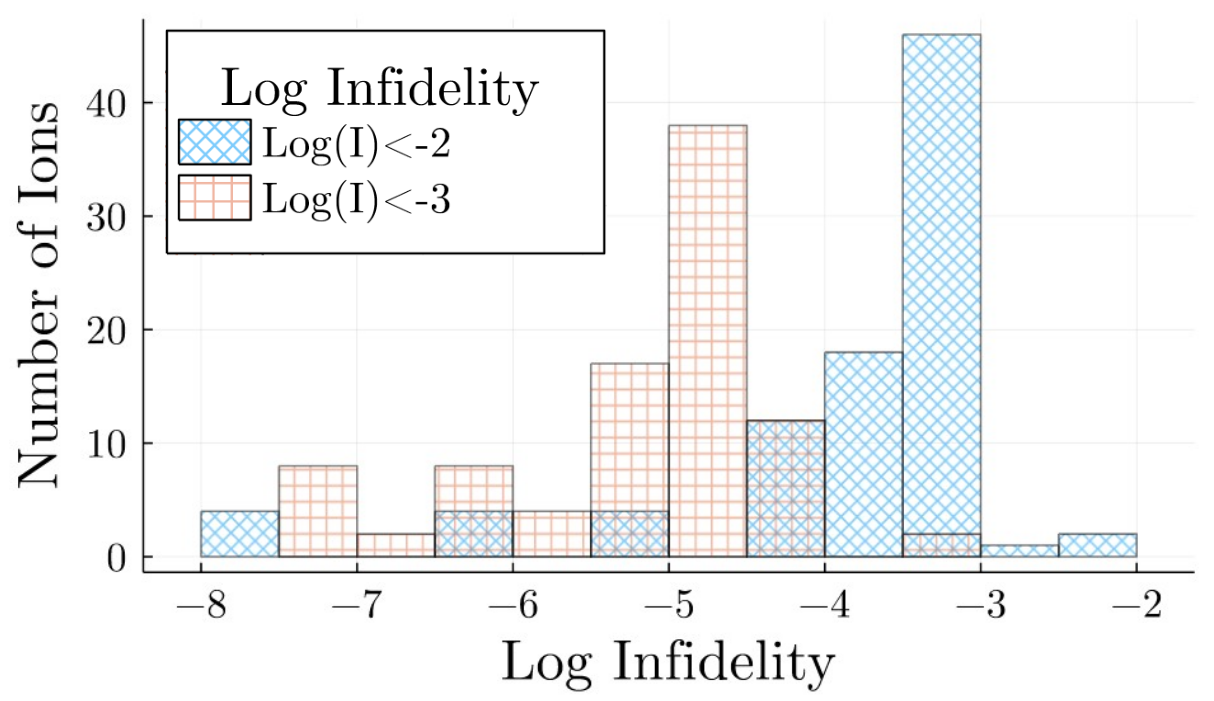}
    \caption{\textbf{}}
    \label{fig:annulus_hist}
    \end{subfigure}
    \captionsetup{justification=raggedright,singlelinecheck=false}
        \caption{Infidelity  $I$ (see Eq. \ref{eq:infidelity}) for preparing an annulus following the protocol of Sec.~\ref{sec:ann_sec}. (a) For $n\leq 24$ and $m=0$, the  maximum infidelity is smaller than $10^{-2}$.(b) For $n\leq 54$ and $m=0$, the maximum infidelity is  smaller than $10^{-3}$. (c) A histogram over the ions' infidelities, comparing (a) and (b).}
    \label{fig:annulus_infid}
\end{figure*}

The expectation values $\braket{\sigma_X}$ after performing the precompensation protocol is shown in Fig.~\ref{fig:annulus_expectation}, and the infidelity to the target state is shown in Fig.~\ref{fig:annulus_infid}. Here the infidelity is due to finite $n$ truncation. Figure~\ref{fig:annulus_infid1} shows an infidelity better than $10^{-2}$ for $n_{\rm max}=24$. An infidelity of $10^{-3}$ can be obtained with $n_{\rm max} = 54$ as shown in Fig.~\ref{fig:annulus_infid2}. A histogram of the infidelities for the two different $n_{\rm max}$ values is shown in  Fig.~\ref{fig:annulus_hist}. The presented analysis shows that the protocol can produce a faithful reconstruction of the annulus pattern with a small state infidelity ($<10^{-3}$).

\subsection{Elliptical Gaussian}\label{sec:egauss}
The next example we consider is an elliptical Gaussian,
\begin{equation}\label{eq:elliptical_Gaussian}
    F(\rho, \phi) = \frac{A}{2}\exp(-(\rho\cos{(\phi)})^2/(2\eta_x^2) - (\rho\sin{(\phi)})^2/(2\eta_y^2)
\end{equation}
We choose $\eta_x=\sqrt{2}/10$ and $\eta_y = \sqrt{2}$ to generate an elliptical pattern with a ``narrow" and ``wide" distribution  covering  the $91$-ion crystal. Because there is no longer azimuthal symmetry, we must consider $m>0$.

\subsubsection{Reconstruction}
\begin{figure}
    \centering
    \includegraphics[scale=.25]{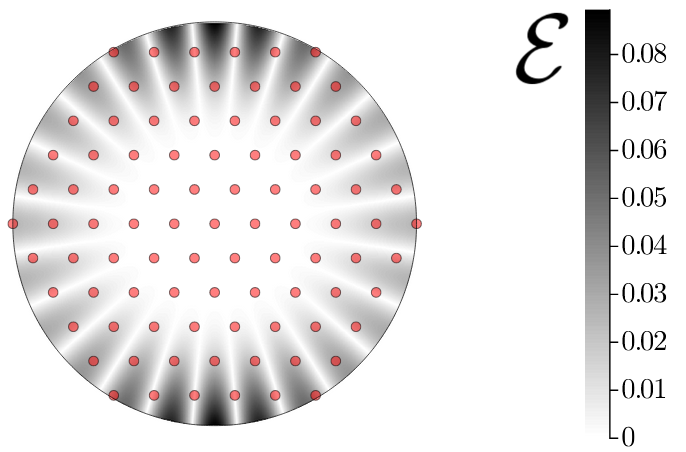}
    \captionsetup{justification=raggedright,singlelinecheck=false}
    \caption{Absolute value of the error $\mathcal{E}=|F - \tilde{F}|/A$ in reconstructing an elliptical Gaussian. By using $n_{\rm max}=26$ and $ m_{\rm max} = 10$  with $m$ always positive we can reconstruct an elliptical Gaussian pattern with a maximum error of less than $0.09$. Nevertheless the maximum error at an ion is approximately $0.035$ since there are no ions sitting at the maximum of the phase pattern. The periodicity in the error pattern is mainly  set by the $m_{\rm max}$ value considered.}
    \label{fig:elliptical_error}
\end{figure}

We use a numerical integrator to compute the Zernike basis coefficients in Eq.~\ref{eq:exact_coefficients}. The reconstruction error for $m_{\rm max}=10$ and $n_{\rm max}=26$ is shown in Fig.~\ref{fig:elliptical_error}. Interestingly, there are radial bands, set by the value of $m_{\rm max}$. For $m_{\rm max}=10$, the next higher-order omitted term has $12$ full periods of oscillation corresponding to the $12$ pairs of dark fringes in Fig.~\ref{fig:elliptical_error}. Since the error is minimized at the bright fringes, we can in principle achieve a higher fidelity reconstruction by adjusting $m_{\rm max}$ so that ions near the edge of the crystal in the $y$-direction are positioned near the minima of the reconstruction error. At any ion in the crystal, the reconstruction error $\mathcal{E}$ shown in Fig.~\ref{fig:elliptical_error} is less than $0.035$, which should enable an implementation of the elliptical Gaussian AC Stark shift pattern with an infidelity less than $10^{-2}$.

The error in the reconstruction is seen to be smaller near the center of the disk, since including lower-degree radial polynomials can match the behavior there. To capture the phase pattern far from the center of the disk requires higher order radial polynomials. In fact the error is observed to be maximum at the vertical wings of the distribution since to capture the decay of the Gaussian  closer to the $\rho=1$ boundary requires higher order terms.

\subsubsection{Series Application}
\begin{figure}
    \centering
    \includegraphics[scale=.25]{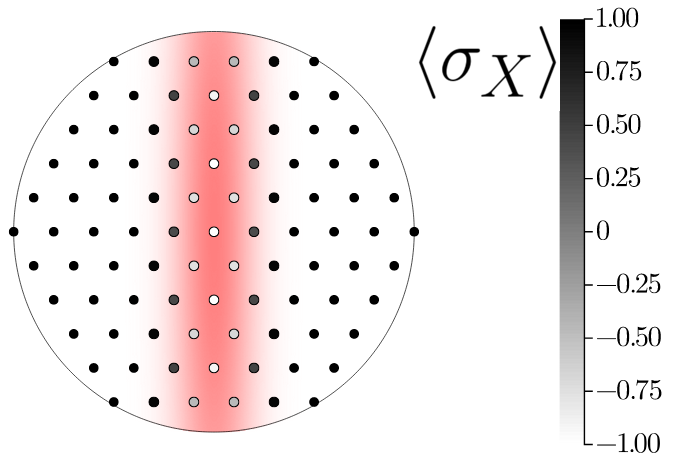}
    \captionsetup{justification=raggedright,singlelinecheck=false}
    \caption{$\langle\sigma_X\rangle $ for an elliptical Gaussian. We use $n\leq 26$, $0 \leq m \leq 10$, and $A=0.5$. Ions along the center of the elliptical phase pattern  are rotated towards $\ket{-}$, while those far away from the center remain in $\ket{+}$.}
    \label{fig:elliptical_expectation}
\end{figure}

\begin{figure*}
  \begin{subfigure}[t]{0.3\textwidth}
    \centering
    \includegraphics[scale=.17]{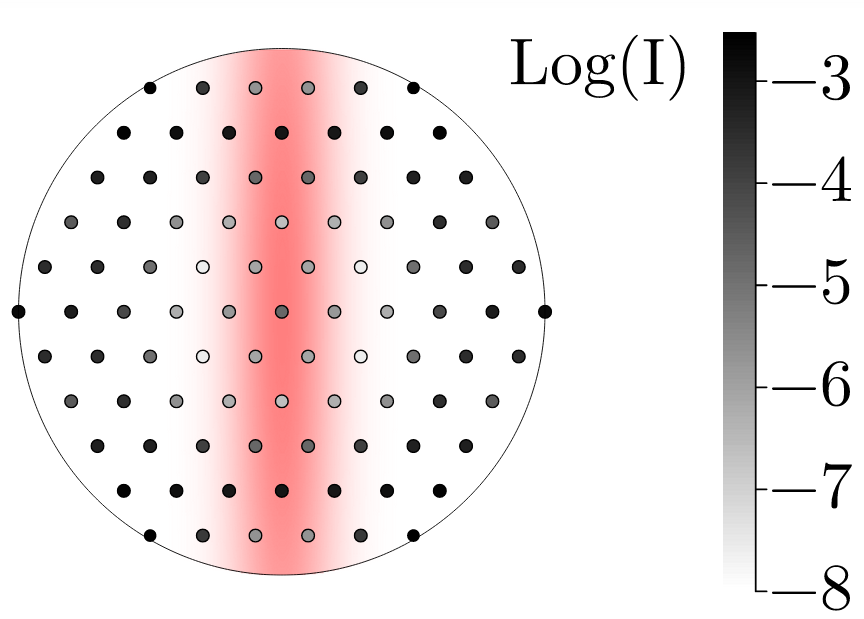}
    \caption{\textbf{}}
    \label{fig:elliptical_infid1}
    \end{subfigure}
    \begin{subfigure}[t]{0.3\textwidth}
    \centering
    \includegraphics[scale=.17]{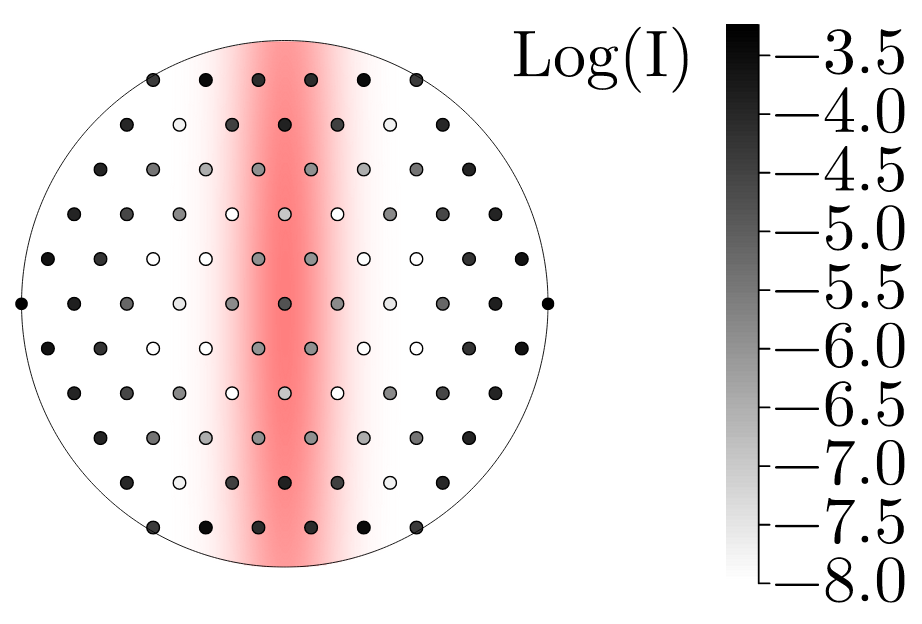}
    \caption{\textbf{}}
    \label{fig:elliptical_infid2}
    \end{subfigure}
    \begin{subfigure}[t]{0.3\textwidth}
    \centering
    \includegraphics[scale=.16]{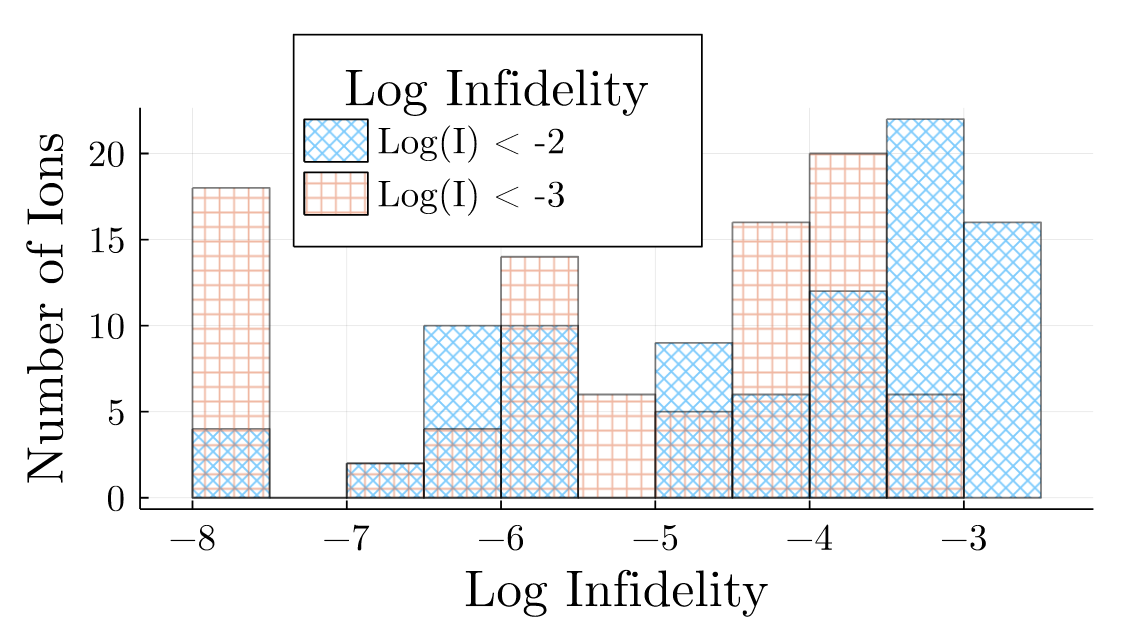}
    \caption{}
    \label{fig:elliptical_hist}
    \end{subfigure}
    \captionsetup{justification=raggedright,singlelinecheck=false}
    \caption{Infidelity for preparing an elliptical Gaussian with the serial protocol. (a) For $n\leq 26$, $0 \leq m \leq 10$, and $A=0.5$, the maximum infidelity is smaller  than $10^{-2}$. (b) For $n\leq 32$, $0 \leq m \leq 12$, and $A=0.5$, the  maximum infidelity is smaller than $10^{-3}$. (c) A histogram over the ions' infidelities, comparing (a) and (b).}
    \label{fig:elliptical_infidelity}
\end{figure*}

Figure.~\ref{fig:elliptical_expectation} shows $\langle\sigma_X\rangle$ at the end of a series evolution with $m_{\rm max}=10$, $n_{\rm max}=26$, and $A=0.5$. Note that because of the elliptical symmetry, $m$ is restricted to non-negative even values, corresponding to six terms in total. We choose a gate time of $T=18\times (2\pi/\omega)=100\,\mu s$ and hence we  remove the  errors from the rotating wave approximation. The protocol generates spin rotations along the three vertical columns of ions in the center of the disk, while very nearly leaving all other ions in the $\ket{+}$ state. Figure~\ref{fig:elliptical_infid1} shows a maximum single-spin infidelity of $10^{-2}$.  With $m_{\rm max}=12$,  corresponding to seven terms, and   making  $n_{\rm max}=32$, Fig.~\ref{fig:elliptical_hist} shows clearly that the maximum infidelity is reduced to $10^{-3}$.  

Each application of one even ($AP^m(\rho) \cos{ (m\phi)}$) term in Eq.~\ref{eq:phase_rewrite} takes $100~\mu$s.  When done in series, the $6$ different values of $m$ employed to obtain an infidelity of $10^{-2}$ sets a gate time of $600~\mu$s, neglecting the reset time of the DM. A gate time of approximately $700~\mu$s is required for the seven terms that enable an infidelity of $10^{-3}$. The reset time of the DM can significantly increase the gate time.  For example, a DM reset time of $50~\mu$s increases the gate time by $50\%$.

\subsubsection{Parallel Application}\label{sec:pegauss}
\begin{figure*}
  \begin{subfigure}[t]{0.3\textwidth}
    \centering
    \includegraphics[scale=.17]{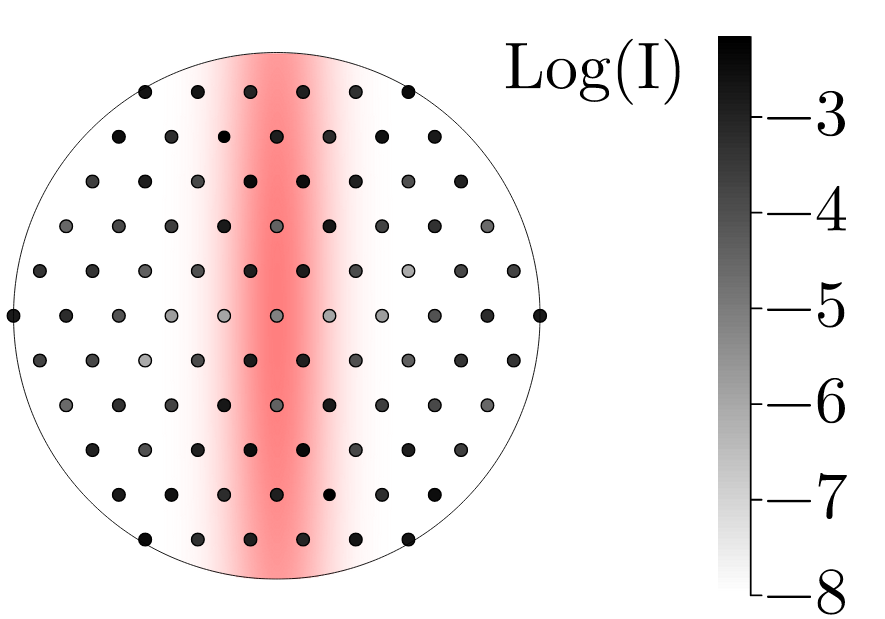}
    \caption{\textbf{}}
    \label{fig:pelliptical_infid1}
    \end{subfigure}
    \begin{subfigure}[t]{0.3\textwidth}
    \centering
    \includegraphics[scale=.17]{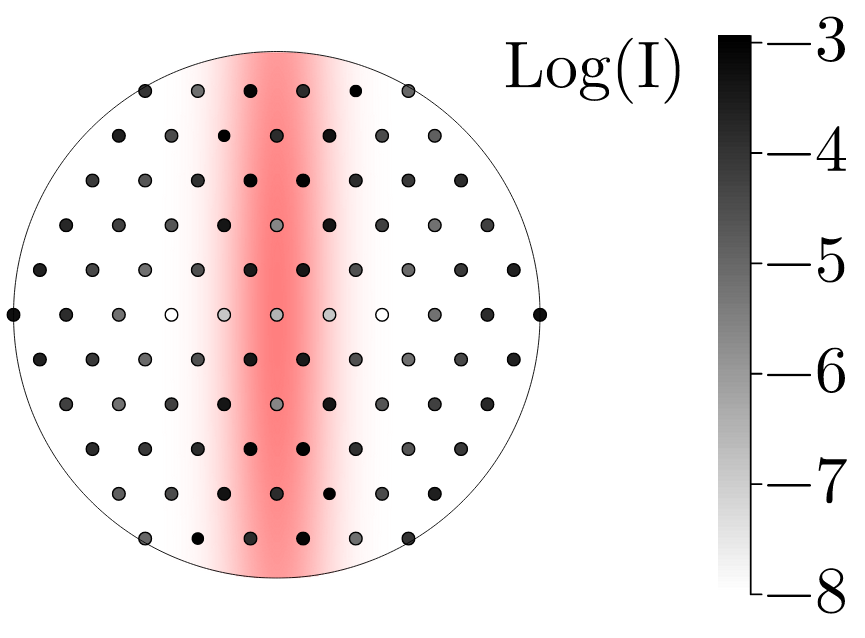}
    \caption{\textbf{}}
    \label{fig:pelliptical_infid2}
    \end{subfigure}
    \begin{subfigure}[t]{0.3\textwidth}
    \centering
    \includegraphics[scale=.16]{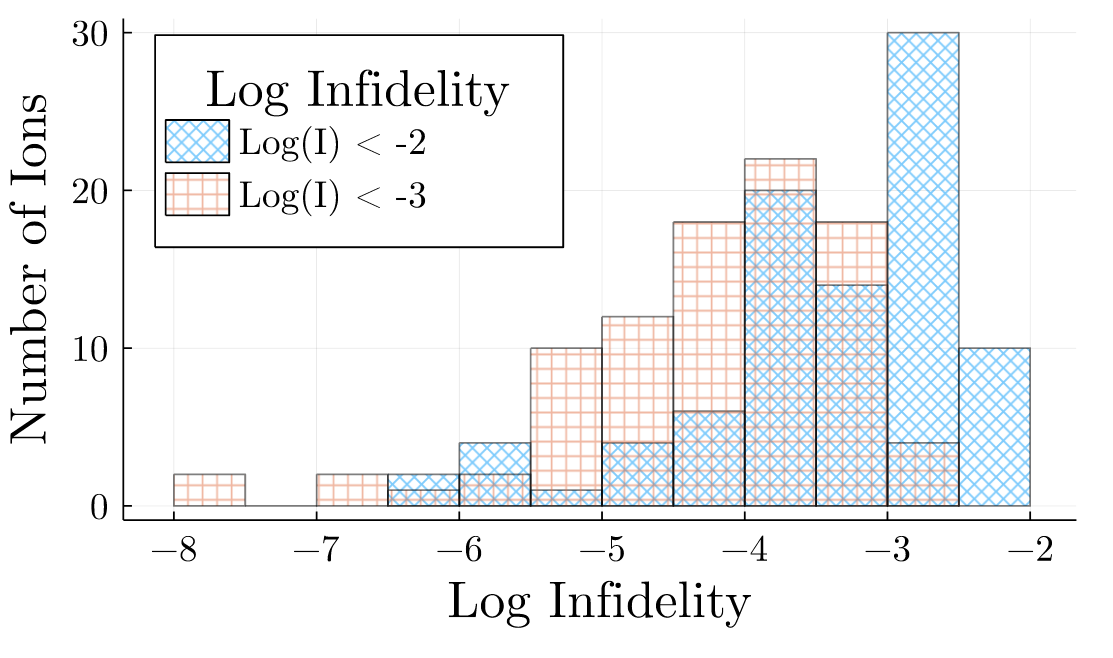}
    \caption{}
    \label{fig:pelliptical_hist}
    \end{subfigure}
    \captionsetup{justification=raggedright,singlelinecheck=false}
    \caption{Infidelity for preparing an elliptical Gaussian with the parallel protocol. (a) For  $n\leq 26$ and $0 \leq m \leq 10$. By applying all polynomials and beatnotes simultaneously, with $A = 0.4$, the resulting  maximum infidelity is less than $10^{-2}$. (b) With $n\leq 32$, $0 \leq m \leq 12$, but and with $A = 0.2$, the maximum infidelity is no more than $3\times10^{-3}$ with all polynomials and beatnotes applied simultaneously. (c) A histogram over the ions' infidelities, comparing (a) and (b). 
    }
    \label{fig:pelliptical}
\end{figure*}

The protocol where all terms are applied in parallel has the advantage that the DM is only set once. This can possibly lead to shorter gate times. The targeted AC Stark shift (Eq.~\ref{eq:elliptical_Gaussian}) and choices of parameters for the parallel application are the same as those considered in the previous paragraphs with the exception of the choice of $A$ and therefore $T$. The parameter $A$ is chosen sufficiently small so that the linear approximation discussed in Sec.~\ref{sec:setup} holds, and $T$ is picked so that the spins at the maxima of the phase pattern experience a full $\pi$ rotation, as discussed in Sec.~\ref{sec:simulation}. By applying the different orders in parallel, we incur all of the errors from the previous section and additional errors from the linear approximation discussed in Sec.~\ref{sec:linear-approx}. Because the $n$ and $m$ chosen in the previous section were minimally large to meet our fidelity requirements, we keep them the same. For a maximum infidelity of $10^{-2}$, Fig.~\ref{fig:pelliptical_infid1} demonstrates that choosing $A=0.4$ is sufficient. This changes the gate time to $250~\mu$s, obtained with $T = 45\times (2\pi/\omega)$. If we increase our infidelity requirements to $3\times10^{-3}$, we see in Figs.~\ref{fig:pelliptical_infid2} and \ref{fig:pelliptical_hist} that we can choose $A=0.2$. For the gate time to be commensurate with the crystal rotation frequency we find $T = (90 \times 2\pi/\omega)$ giving a longer gate time of $500~\mu$s. We note that we have chosen $3\times10^{-3}$ as the target infidelity rather than $10^{-3}$ as in the other examples that we show. 

\subsection{Displaced Gaussian}\label{sec:edisplaced}
Finally, we consider an AC Stark shift pattern described by a displaced Gaussian,
\begin{widetext}
\begin{equation}
    F(\rho, \phi) = \frac{A}{2}\exp(-((\rho\cos{(\phi)}-\delta_x)^2 + (\rho\sin{(\phi)}-\delta_y)^2)/(2\eta^2))
\end{equation}
\end{widetext}
with standard deviation $\eta = 0.1/\sqrt{2}$, displaced by $\delta_x = 0.3$ in $x$ and $\delta_y = 0.1\sqrt{3}$ in $y$. The displacement was chosen to coincide with an ion in the crystal, and the width was chosen to achieve a  single spin rotation. In particular, this choice of $\eta$ corresponds to a Gaussian profile that decays by a factor of $1/e$ at a diameter of the inter-particle spacing. In this case, we no longer have azimuthal symmetry, so we must include $m \geq 0$ and $m < 0$ as well.
\subsubsection{Reconstruction}
\begin{figure}
    \centering
    \includegraphics[scale=.25]{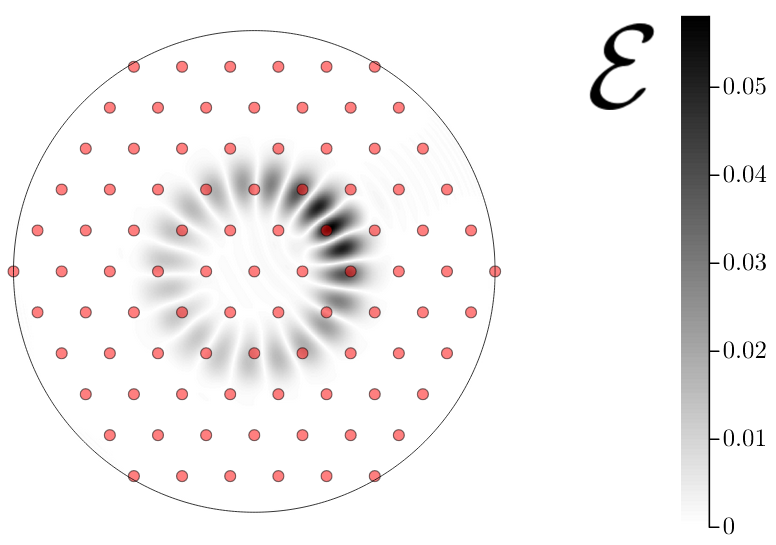}
    \caption{Absolute value of the error $\mathcal{E}=|F - \tilde{F}|/A$ in reconstructing a displaced Gaussian. With $n_{\rm max}=40$ and $|m_{\rm max}|= 9$ we get a maximum error smaller than $0.06$. The error is worse around the targeted ion since this is the sharpest feature trying to be reconstructed with $20$ dark fringes corresponding to the extrema of the next omitted $m=10$ term.}
    \label{fig:displaced_error}
\end{figure}

In Fig.~\ref{fig:displaced_error}, we included  up to $m=\pm 9$ and $n_{\rm max} = 40$ terms. We see that the reconstruction has a maximal error in a region surrounding the targeted ion. The maximum truncation error is approximately $0.06$, which should enable a single-spin infidelity of $10^{-2}$. We note that choosing an ion closer to the boundary of the crystal will require a large  $m$ value to reach similar truncation error.

\subsubsection{Series Application}
\begin{figure}
    \centering
    \includegraphics[scale=.25]{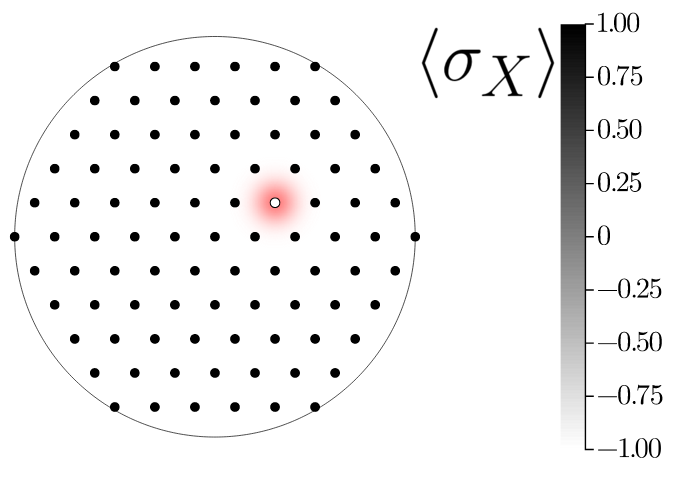}
    \caption{$\langle\sigma_X\rangle $  for a displaced Gaussian after following the protocol in Sec.~\ref{sec:simulation}. Using $n_{\rm max}=40$ and $|m_{\rm max}| = 9$, we see that a single ion spin is rotated to a very good approximation.}
    \label{fig:displaced_expectation}
\end{figure}

\begin{figure*}
  \begin{subfigure}[t]{0.3\textwidth}
    \centering
    \includegraphics[scale=.17]{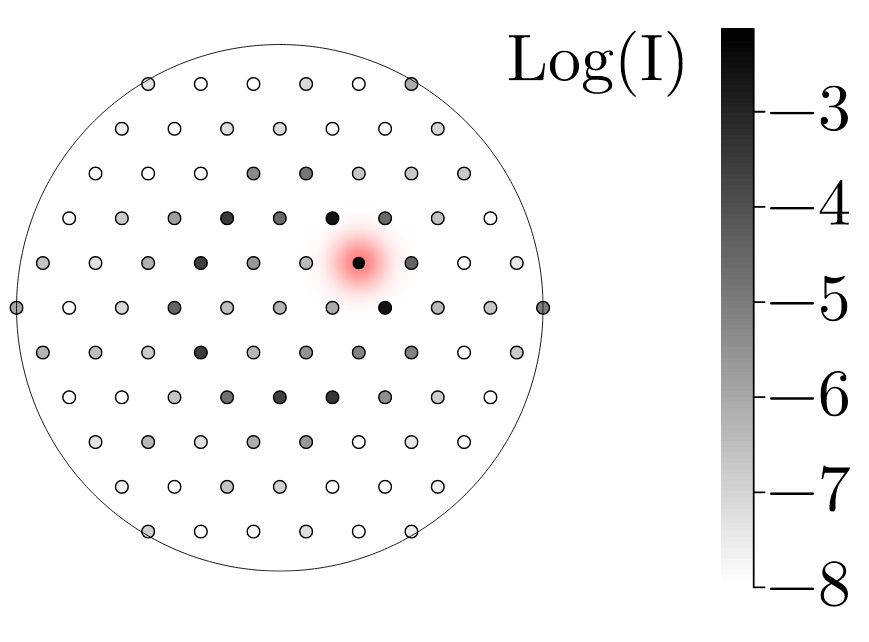}
    \caption{\textbf{}}
    \label{fig:displaced_infid1}
    \end{subfigure}
    \begin{subfigure}[t]{0.3\textwidth}
    \centering
    \includegraphics[scale=.17]{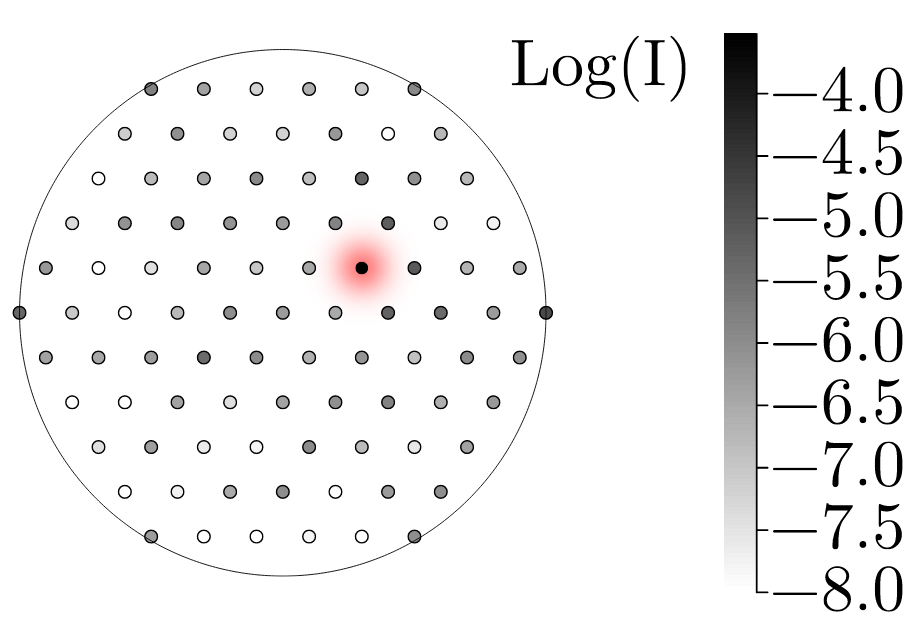}
    \caption{\textbf{}}
    \label{fig:displaced_infid2}
    \end{subfigure}
    \begin{subfigure}[t]{0.3\textwidth}
    \centering
    \includegraphics[scale=.16]{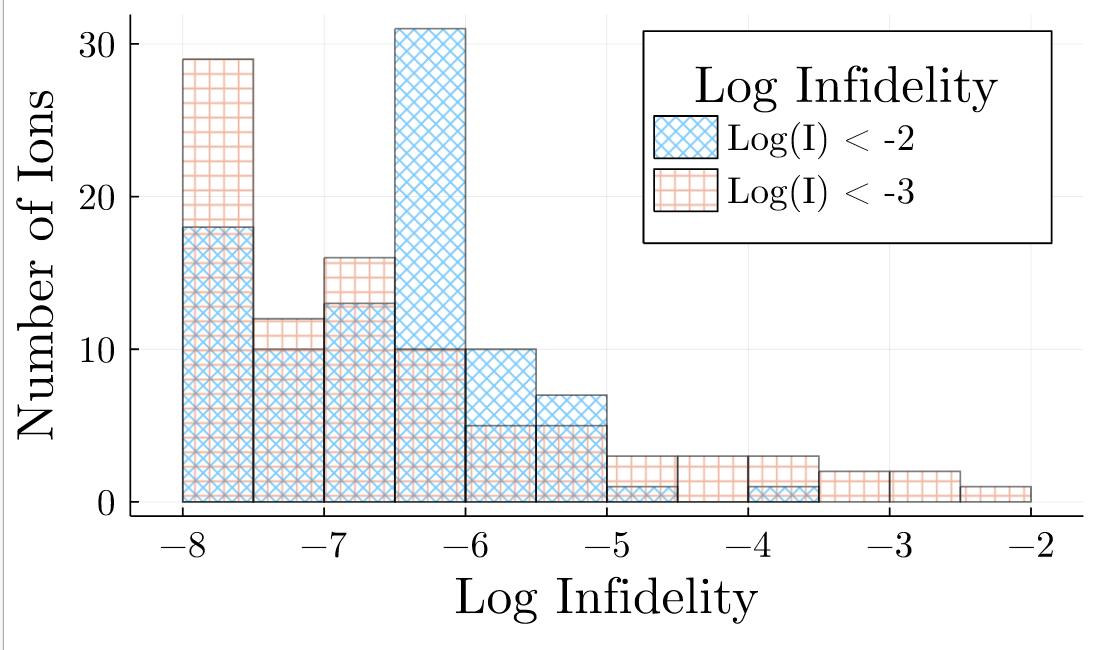}
    \caption{}
    \label{fig:displaced_hist}
    \end{subfigure}
    \caption{Infidelity for preparing a displaced Gaussian with the serial protocol. (a) For $n\leq 40$, $-9 \leq m \leq 9$ and $A=3.0$ the maximum infidelity is smaller than $10^{-2}$. (b) For $n\leq 40$, $-20 \leq m \leq 20$ and $A=3.0$ the maximum infidelity is smaller than $10^{-3}$. (c) A histogram over the ions' infidelities, comparing (a) and (b). }
    \label{fig:displaced_gaussian}
\end{figure*}

In this example we consider $A=3.0$, motivated by our infidelity goal of $10^{-2}$. We have picked this value of $A$ to be as large as possible while still requiring that each term being applied in the expansion (Eq.~\ref{eq:phase_rewrite}) can be inverted as discussed in Sec.~\ref{sec:setup} for precompensation. Additionally, we have chosen  $T=3\times(2\pi/\omega)$, to remove the error from the RWA. For $A=3.0$, the application time required for each beatnote is approximately $16.66~\mu$s. When the polynomials are applied in series for $19$ different values of $m$, we get a gate time of approximately $316.66~\mu$s, again assuming a DM with zero reset time. If we increase our infidelity goals to $10^{-3}$, we can choose $|m| \leq 20$ and $A=3.0$, giving a gate time of $683.33~\mu$s. As in the previous two cases, we see excellent agreement with the desired phase pattern in the evolution of $\langle\sigma_X\rangle$ shown in  Fig.~\ref{fig:displaced_expectation}.
The protocol very nearly rotates a single spin, as all of the ions surrounding the desired ion are very nearly in the $\ket{+}$ state. This behavior leads us to conclude that single ion addressability is feasible with this technique, requiring about a factor of two more terms than in the other cases, due to the additional odd ($AQ^m\sin{(m\phi)}$) terms. The infidelities in Fig.~\ref{fig:displaced_infid1} and Fig.~\ref{fig:displaced_infid2} are worst in a radial band of ions containing the ion being flipped. This is intuitive - suppressing that error requires the introduction of high order angular terms, while we have used only up to $m=9$. 

\subsubsection{Parallel Application}
\begin{figure*}
  \begin{subfigure}[t]{0.3\textwidth}
    \centering
    \includegraphics[scale=.18]{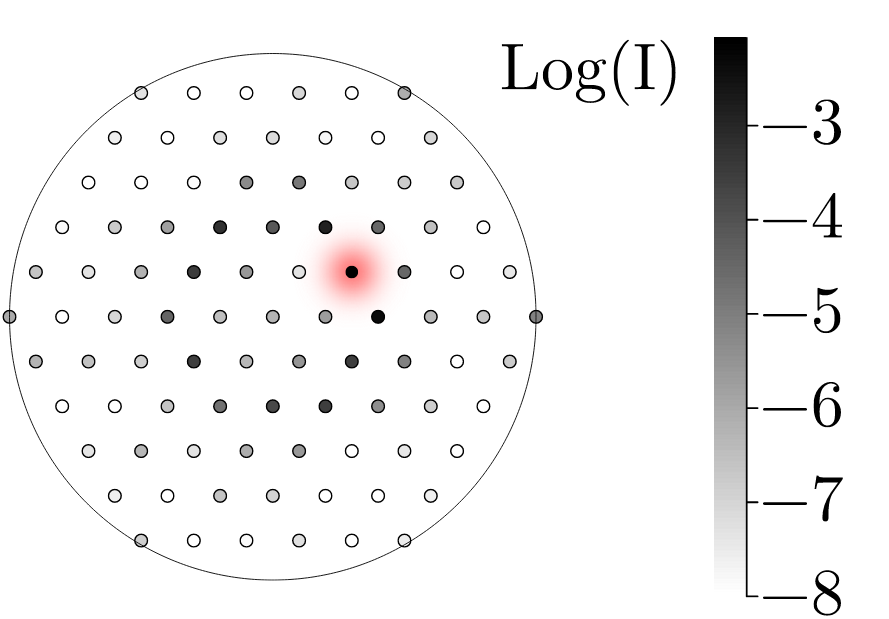}
    \caption{\textbf{}}
    \label{fig:pdisplaced_infid1}
    \end{subfigure}
    \begin{subfigure}[t]{0.3\textwidth}
    \centering
    \includegraphics[scale=.17]{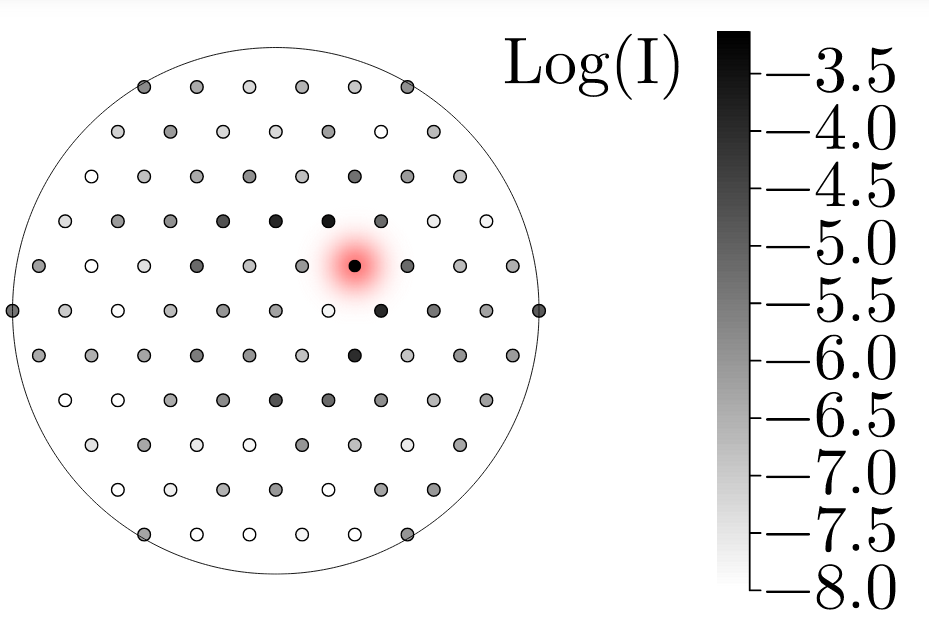}
    \caption{\textbf{}}
    \label{fig:pdisplaced_infid2}
    \end{subfigure}
    \begin{subfigure}[t]{0.3\textwidth}
    \centering
    \includegraphics[scale=.16]{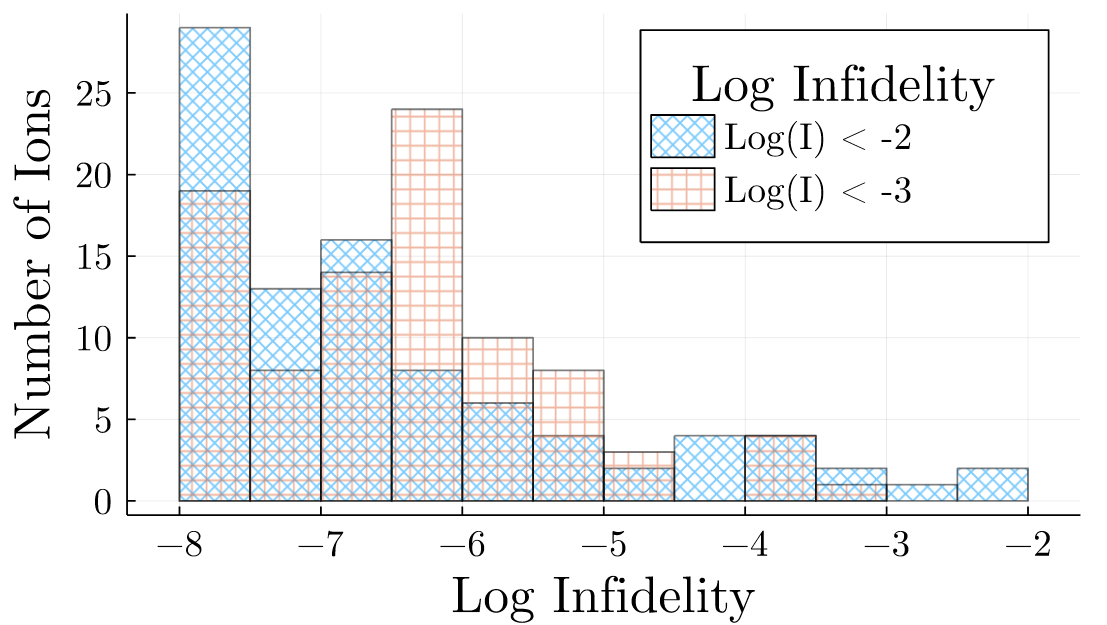}
    \caption{}
    \label{fig:pdisplaced_hist}
    \end{subfigure}
    \caption{Infidelity for preparing a displaced Gaussian with the parallel protocol. (a) For $n\leq 40$ and $-9 \leq m \leq 9$. By applying all polynomials and beatnotes simultaneously, with $A=0.3$, the maximum infidelity is smaller than $10^{-2}$. (b) For $n\leq 40$, $-20 \leq m \leq 20$, and still with $A=0.3$, the maximum infidelity is less than $10^{-3}$. (c) A histogram over the ions' infidelities, comparing (a) and (b).}
    \label{fig:pdisplaced}
\end{figure*}

The parameters in this section are the same as those considered in the previous paragraphs, with the exception of the choice of $A$ and therefore $T$. Figure~\ref{fig:pdisplaced_infid1} and Fig.~\ref{fig:pdisplaced_infid2} demonstrate that choosing $A=0.3$ is sufficient for the  $10^{-2}$ and $10^{-3}$ infidelity requirements. Consequently, the gate time in both cases is approximately $333.33~\mu$s, which is given as $T=60\times(2\pi/\omega)$ and thus the error incurred from the rotating wave approximation is zero. This reduces the gate time for the parallel application compared to the serial application only for the case of a targeted infidelity of $10^{-3}$. However, this neglects the reset time of the DM. For a targeted infidelity of $10^{-2}$, the serial application needed $m_{\rm max} = 9$, which would require the DM to be set $18$ times. With a reset time as high as $50~\mu$s, this incurs an overhead of $900~\mu$s. This is substantially larger than the gate time itself, and highlights a potential reason to instead consider the parallel protocol.

\section{Conclusion}
Penning traps are promising candidates as platforms for quantum information processing, due to their ability to control hundreds of qubits, and perform non-local entangling operations. However, existing experiments lack the ability to address individual ions, and therefore fail to meet the criteria for universal quantum information processing. In this paper we have discussed a method for implementing programmable $Z$-rotations in a Penning trap, thus providing a path forward for more complex quantum simulations and general large-scale quantum information processing.

By employing a deformable mirror (DM) in the path of one of the laser beams that creates the optical dipole force in Refs.~\cite{britton2012engineered,Bohnet2016}, we showed how wave front deformations introduced by the DM can be used to generate AC Stark shift patterns that are static in the rotating frame of the crystal.  A pattern of azimuthal order $m$ ($P^m(\rho)\cos (m\phi), Q^m(\rho)\sin (m\phi)$) is introduced by setting the frequency $\mu$ of the optical dipole force to the $m^{th}$ harmonic of the rotation frequency, $\mu = m\omega$.  General AC Stark shift patterns are obtained through the introduction of different azimuthal orders.  We analytically and numerically demonstrated the feasibility of this approach for generating single-site rotations.  Choosing a Gaussian phase pattern with a 1/e diameter equal to the interparticle spacing, we demonstrated that we can resolve single ions in a crystal of about 100 ions under typical experimental conditions \cite{britton2012engineered,Bohnet2016}.  Moreover, by applying the required beatnote frequencies $m\omega$ in parallel, one can obtain $99\%$ fidelity single-qubit gate times of $333.33~\mu$s, which is faster than typical single particle decoherence times, $\sim 10$ ms, in current Penning trap experiments.

In this manuscript we assume a perfectly performing DM and analyze the sources of infidelity for two different protocols where patterns of different azimuthal order are introduced serially or in parallel.   In particular, we assume that the number of DM actuators is large compared to the number of ions and that the surface of the DM can be set with arbitrary precision.  Deformable mirrors with greater than 3000 actuators, surface figures of less than 10 nm, and mechanical response times of less than 50~$\mu$s are available commercially.  The desired AC Stark shift pattern as well as the performance of the DM will impact whether the serial or parallel protocol should be employed.  In general, the serial protocol enables the implementation of larger amplitudes and therefore higher accuracy phase patterns.  However, for patterns that require introducing many azimuthal orders $m$, the reset time of the DM can add significant overhead in the time required to implement the desired AC Stark shift pattern.   The parallel protocol removes any overhead due to the reset time of the DM, but the restriction on the amplitude of the phase pattern can impact the accuracy with which the desired phase pattern can be implemented.

\section{Acknowledgments}
We acknowledge helpful discussions with Christian Marciniak, and thank Allison Carter and Jennifer Lilieholm for reading and commenting on our manuscript. This material is based upon work partially supported by the U.S. Department of Energy, Office of Science, National Quantum Information Science Research Centers, Quantum Systems Accelerator (QSA). AMP acknowledges funding from NSF grant number 1734006 and a NASA Space Technology Graduate Research Opportunity. JJB acknowledges support from the DARPA ONISQ program and AFOSR grant FA9550-20-1-0019. 

\bibliography{refs}

\newpage
\appendix

\clearpage

\onecolumngrid
\section{The RWA is Exact for Integer Multiples of the Trap Rotation Frequency}
As discussed in Sec.~\ref{sec:rwa}, choosing the evolution time $T$ such that $\omega T =   2\pi r$ with $r$ a positive integer can remove all error from the rotating wave approximation. In Sec.~\ref{sec:rwa} we showed that this was the case when we applied different $P_m$ patterns.
We now argue that this can be generalized to arbitrary order. The Hamiltonian we consider will have both even ($AP^{m_i}\cos{(m_i\theta)}$) and odd ($AQ^{m_i}\sin{(m_i\theta)}$) terms, up to $m_{N_f}$, so that $i \leq N_f$. From the Jacobi-Anger expansion each of these terms will introduce a new sum, as in Eq.~\ref{eq:doublesum}, with an index $a_i$ or $b_i$ respectively. In general, we then get phase factors in the sum, $f(t)$, of the form
\begin{equation}
    f(t) = \exp{(-i\mu t -i\phi\sum_i (a_i+b_i)m_i - i\sum_i (a_i+b_i)m_i\omega t)}.\label{ea}
\end{equation}
Because this is the only time dependence, we can integrate over time from $0$ to $2\pi r/\omega$ to get
\begin{equation}
    \int_0^{2\pi k/\omega} dt f(t) = \frac{\exp{(-i\mu t - i\phi\sum_i (a_i+b_i)m_i- i\sum_i (a_i+b_i)m_i\omega t) )}}{-i(\mu + \sum_i (a_i+b_i)m_i\omega)}\bigg\rvert_0^{2\pi r/\omega} = 0
\end{equation}
 This is true except when the denominator vanishes, which is also the case where the term in Eq. \ref{ea} is static.  If we have a drive $\mu = m_j\omega$ the denominator vanishes when
\begin{equation}
    m_j + \sum_{i=1}^{N_f} (a_i+b_i)m_i = 0,
\end{equation}
as we have seen previously in the case of two terms. 
The case in Sec.~\ref{sec:rwa} was special in that it repeats every $\pi/\omega$. The general argument is given above, and shows that in general a multiple of $2\pi/\omega$ is needed.
\end{document}